\documentclass{aa}

\usepackage{graphicx}
\usepackage{txfonts}
\usepackage{amsmath}
\DeclareMathOperator{\asinh}{asinh}
\usepackage{natbib}
\begin{document}

   \title{\emph{Gaia} Data Release 1. Cross-match with external catalogues }

   \subtitle{Algorithm and results}

   \author{P.M. Marrese \inst{1,2}\fnmsep\thanks{ \email{paola.marrese@ssdc.asi.it}}
       \and S. Marinoni \inst{1,2}
       \and M. Fabrizio \inst{1,2}
       \and G. Giuffrida \inst{1,2}
   }
   \institute{Space Science Data Center - ASI, via del Politecnico SNC, I-00133 Roma, Italy
   \and INAF-Osservatorio Astronomico di Roma, Via Frascati 33, I-00040, Monte Porzio Catone (Roma), Italy\\ }
       
   \date{Received ; accepted}

 
  \abstract
  {Although the \emph{Gaia} catalogue on its own will be a very powerful tool, it is the combination of this highly accurate archive with other archives that 
   will truly open up amazing possibilities for astronomical research. The advanced interoperation of archives is based on cross-matching, leaving 
   the user with the feeling of working with one single data archive. The data retrieval should work not only across data archives, but also across
   wavelength domains. 
   The first step for  seamless data access is the computation of the cross-match between \emph{Gaia} and external surveys.}
  {The matching of astronomical catalogues is a complex and challenging problem both scientifically and technologically (especially when matching large surveys like \emph{Gaia}). 
  We describe the cross-match algorithm used to pre-compute the match of  \emph{Gaia} Data Release 1 (DR1)  with a selected list of large publicly available optical and IR surveys. }
  {The overall principles of the adopted cross-match algorithm are outlined. Details are given on the developed algorithm, including the methods used to account for position 
  errors, proper motions, and environment; to define the neighbours; and to define the figure of merit used to select the most probable counterpart.} 
  {Statistics on the results are also given. The results of the cross-match are part of the official \emph{Gaia} DR1 catalogue.}
   {}


   \maketitle
%
\section{Introduction}\label{intro}
The \emph{Gaia} satellite will allow the  positions,
parallaxes, and proper motions to be determined with high accuracy for more than 1 billion sources reaching magnitude
$G\sim20.7$. All \emph{Gaia} sources will also have multicolour photometry, while radial velocities will only be available for sources brighter than $G\sim17$. The summary of the astrometric, photometric, and survey properties of \emph{Gaia} Data Release 1 (DR1) are described in \citet{Brown2016}, while the scientific goals of the mission are summarised in  \citet{Prusti2016}. 
Combining the \emph{Gaia} catalogue with other publicly available surveys obtained either from ground or from space   more closely meets the 
modern astronomical research requirements.
The main aim of  adding a pre-computed cross-match to the official \emph{Gaia} DR1 data is to complement \emph{Gaia} with existing (and widely used by the scientific community) photometry and astrometry, thus allowing the full scientific exploitation of \emph{Gaia}. 
The complexity and  scientific issues related to cross-matching has become very popular now that the combined use of large data sets from different surveys 
and/or wavelength domains is more and more common.
 \citealt{arenou}  shows how the comparison with external catalogues allows a deeper understanding of many of the parameters describing the performances
of the \emph{Gaia} catalogue. The results of the cross-match described here played an important role in the full sky tests utilised for the validation 
of \emph{Gaia} DR1 data, constituting the first scientific exploitation of the cross-match results described in this paper.

In the following, detailed explanations of the general principles we followed and of the reasons behind our choices on each and all the scientific issues 
of the adopted cross-match algorithm is given. A detailed pinpointing of the caveats and of the failed cases is also available, allowing the scientists who 
 use the cross-match results to be fully aware both of  its quality and of its possible limitations.

The cross-matching (hereafter XM)  of astronomical catalogues is a complex and challenging problem both scientifically and technologically, especially 
when matching large surveys which include several millions or billions of sources. 
In this paper we concentrate on the scientific issues, thus only a short description of  the technological and computational implementation is given.

There are different approaches to the XM of astronomical catalogues, and XM algorithms can also be very different.
It is important to correctly define both the scientific problem one is faced with and the objectives of the cross-match. 

When a neighbour in the secondary catalogue is found close to a leading catalogue source, 
the first question to be answered is whether it should always be considered as the actual counterpart or not.
In the second case the algorithm gets more complicated and some kind of {a priori} knowledge on the 
nature of the object that is being matched becomes necessary. 
The more the two catalogues being matched are different,  the more 
caution should be used in considering the neighbours as counterparts. For example, one may want to match a large general purpose 
survey with a survey of a particular class of objects, or the two catalogues could be largely not homogeneous because they were 
observed in different wavelength domains. 

Depending on the scientific problem, an XM algorithm could require a one-to-one match or allow for a one-to-many or many-to-one matches. 
This will affect the possibility of using a symmetric algorithm.

An XM algorithm is always a trade-off between multiple requisites and a fraction of mismatched objects is always present. 
The scope of a given XM algorithm could be to minimise the absolute number of mismatches, or to minimise the number of 
mismatches among the rarest and most peculiar objects. For example, when matching general purpose surveys, it is important whether to use  
the magnitudes in the selection of the best match. The use of magnitudes and colours requires transformations between photometric systems
that are usually based on synthetic photometry of {normal} stars. While using the magnitudes would  help matching most of the 
objects in a given catalogue, it would probably worsen the matching of many relatively rare but very interesting objects such as variables, 
peculiar stars, and non-stellar objects. In addition, in many surveys not all objects  have a colour (i.e. a fraction of objects may have been 
detected in one band only).
One could aim for a simple algorithm which could be easily applied to many different catalogue pairs or, on the contrary, one could try to tailor
the best possible algorithm for a given scientific problem.
In some cases it would be more important for the same algorithm to be homogeneously applied to all the objects, in others a different definition
of the figure of merit could be allowed (for example, in cases when some {a priori\/} information is available for only a fraction of the objects in a given catalogue).

The scientific details  of a XM  algorithm, in particular which object characteristics (available in the catalogues) to use in the definition of the best  
match,  depend on the pair of catalogues being matched.  The characteristics of each catalogue 
and how the two catalogues compare considering those characteristics are both important.
A  non-exhaustive list of the information that could be used in the definition of an XM algorithm includes the following:
\emph{a)\/} data available (positions, epochs, proper motions, parallaxes, photometry, binary, and/or variability characterisation);
\emph{b)\/} statistics on accuracy and precision of the data available;
\emph{c)\/} photometric depth (magnitude limit) and completeness;
\emph{d)\/} possible systematic errors on any of the data (astrometry and/or photometry) used in the cross-match;
\emph{e)\/} statistics on the availability of the  information within a catalogue (for example how many objects  have colour information);
\emph{f)\/} accuracy of photometric transformations between the two catalogues and their applicability limits; and
\emph{g)\/} angular resolution of each catalogue and the resolution difference between the two catalogues. 

In Section~\ref{sec:gen} we outline the general principles that guided us in the definition of the XM algorithm, while in Section~\ref{sec:details}
we describe the details. In Section~\ref{sec:tech} a brief technical description of the XM implementation is given. Section~\ref{sec:extcat} is dedicated to the description of the external surveys matched with \emph{Gaia}. In Section~\ref{sec:results} we discuss in general terms the XM computation results. Finally in Appendix~\ref{sec:val} a validation test of the method applied to sources with unknown proper motion is discussed.

\section{\emph{Gaia} pre-computed cross-match: general principles\label{sec:gen}}
Among all the different approaches to cross-matching,  we decided to define our algorithm according to the specific scientific problem we have. 

The external catalogues to be matched with \emph{Gaia} are all obtained in the optical/near-IR wavelength region (with the exception of allWISE, which extends in the medium-IR domain), are general surveys not restricted to a specific class of objects and have an angular resolution lower than \emph{Gaia}. As such they are sufficiently homogeneous among themselves to allow the use of a single XM algorithm, which is adapted to each different catalogue using a small number of configurable parameters. 
Since the external catalogues are available together with  \emph{Gaia} DR1 data and their cross-match is part of the official  \emph{Gaia} DR1, consistency and homogeneity in the cross-match computations are an important requirement. We decided to match  \emph{Gaia} DR1 with each external survey separately and independently. A different approach, performing a simultaneous multicatalogue and multiwavelength cross-match (\citealt{Pineau2017}, \citealt{Salvato2017}), is less appropriate in our case as we concentrated our work on large optical/near-IR surveys. 

The algorithm we defined to match \emph{Gaia} data with publicly available 
astrometric/photometric surveys is not symmetric, and we always use \emph{Gaia} as the leading catalogue.
We assume that when a {good neighbour}\footnote{A good neighbour for a given \emph{Gaia} object is a nearby object in the external catalogue whose 
position is compatible within position errors with the target.} is found for a given \emph{Gaia} object, then it is the counterpart. When more than one good neighbour is found, the {best neighbour} (i.e. the most probable counterpart according to the figure of merit we define, see Section~\ref{sec:details}) is chosen from good neighbours.
The higher \emph{Gaia} angular resolution with respect to the external catalogues requires a many-to-one algorithm; this is why the algorithm we used is not symmetric and 
why more than one \emph{Gaia} object can have the same best neighbour in a given external catalogue.
Two or more \emph{Gaia} objects with the same best neighbour are denoted {mates}. True mates are objects resolved by \emph{Gaia}, and are not resolved by the external survey. 

An important requirement of the XM algorithm that we developed is completeness, and we thus defined the position errors to a $5\sigma$ level (see Sect.~\ref{sec:details}.3).
In addition, when defining the XM algorithm, we decided to avoid features which would help the match of generic objects (i.e. normal well-behaved stars) at the cost 
of worsening the match of peculiar classes of objects.
Since we computed the cross-match for several different surveys, we also valued consistency and homogeneity.
We thus decided to avoid the use of a priori knowledge which in general surveys is not usually available for all (or the vast majority of) the objects or for all the external catalogues. 
We tried to avoid  relying too much on assumptions while still using the scientific information present in the input catalogue data. 

The chosen algorithm is positional and thus uses positions, position errors, their correlation if known, and proper motions. 
We used \emph{Gaia} proper motions only, so the proper motion correction was applied only to the TGAS (\emph{Tycho-Gaia} Astrometric Solution) subsample for \emph{Gaia} DR1 ($\sim$2 million objects), while the vast majority of the \emph{Gaia} DR1 stars do not have proper motions.
While the figure of merit we used depends strongly on the angular distance between the \emph{Gaia} target and the external catalogue counterpart candidate and on the position errors, it also  depends on the local surface density of the external catalogue (environment).

We produced two separate XM outputs: a \emph{BestNeighbour} table which lists each  matched \emph{Gaia} object with its best neighbour and 
a \emph{Neighbourhood} table which includes all good neighbours for each matched \emph{Gaia} object (see Section~\ref{sec:results} for a detailed output description).
\subsection{Use of magnitudes in the cross-match algorithm}\label{maginxm}
When  available and depending on the accuracy of the photometric conversion, photometric data can be considered when defining the XM figure of merit.
In order to make use of the magnitudes in the evaluation of the best neighbour and in the process of scoring the neighbours, it is necessary to convert the external 
catalogue magnitudes to the \emph{Gaia} G magnitude. While this is  feasible in principle,  one should bear in mind that in general the transformations between photometric systems  show  quite a large scatter, are not suited for peculiar objects, and have different accuracy for different catalogues. In addition colour information is not generally available for all the objects of a given external catalogue (due to different sensitivity in the different bands) and this in turn causes an inhomogeneous treatment of objects within a given catalogue. We thus avoided  using the photometric information in the best neighbour selection.
\subsection{Use of proper motions in the cross-match algorithm}
Any XM algorithm between two astronomical catalogues is based on object positions and their errors. Proper motions should be taken into 
account when dealing with high proper motion stars or catalogues with very different epochs of observation or in the case of high 
confusion, high density regions.
For the TGAS subsample we moved \emph{Gaia} objects to the individual epoch of the possible matches (i.e. stars in the external 
catalogue which are within the search radius, see Section~\ref{sec:details}) and we propagated their position errors. This approach requires  calculating \emph{Gaia} object 
positions on the fly, rather than computing them to a median external catalogue epoch beforehand.
We decided to discard the possibility of using  the external catalogues' proper motions even when they are available  
as it introduces an inhomogeneity in the XM of  the different catalogues. There is an additional problem when using the external catalogues' proper 
motions: while the catalogues  usually contain positions at a reference epoch (J2000.0), the errors on the positions are given at a mean 
epoch. By definition, the mean epoch is the epoch which minimises the position errors in the proper motion fitting procedure and often the mean epoch is different for 
Right Ascension and Declination. It is of course possible to propagate errors from the mean to the reference epoch, but this implies an approximation 
as the coordinates at the mean epoch are not usually available. 
The algorithm we used to propagate an object position (and position errors) at a different epoch is described in the Hipparcos and Tycho 
Catalogues documentation (\citealt{Hipparcos} first volume, in particular Sects. 1.2 and 1.5
\footnote{The original Hipparcos tool {\sl pos\_prop} implementing the algorithm is also made available (C and FORTRAN) at the following link:\\
\url{http://cdsarc.u-strasbg.fr/viz-bin/Cat?cat=I\%2F239\%2Fversion_cd\%2Fsrc&target=http&}}). The adopted algorithm is based on a standard model of
stellar motion, which assumes that stars move through space with a constant velocity vector. The rigorous treatment of the epoch transformation requires that the 
variation of all six parameters, $\alpha$ (Right Ascension), $\delta$ (Declination), $\pi$ (parallax), $\mu_{\alpha*}$ (proper motion in $\alpha\cos\delta$), 
$\mu_{\delta}$ (proper motion in $\delta$), and $V_{\mathrm{R}}$ (Radial Velocity), must be considered. In our case we used only positions and proper motions; 
we did not use the parallax and radial velocity. As stated in \citealt{Hipparcos} first volume, the simple 
formula\footnote{$\alpha=\alpha_{0} + (T - T_{0})\mu_{\alpha*0}\sec{\delta}$;   ~$\delta = \delta_{0} + (T - T_{0})\mu_{\delta0}$ } 
for transforming a celestial position from one epoch to a different one
is not a good physical model for the star motion. The difference with respect to a rigorous model may become significant near the celestial poles or when propagating the position over a long time.
Including proper motions does not address the astrometric binary problem as it would be necessary to include  the binary orbits when available. 
\subsection{Accounting for epoch differences}
\begin{figure}[]
\centering
\includegraphics[width=90mm]{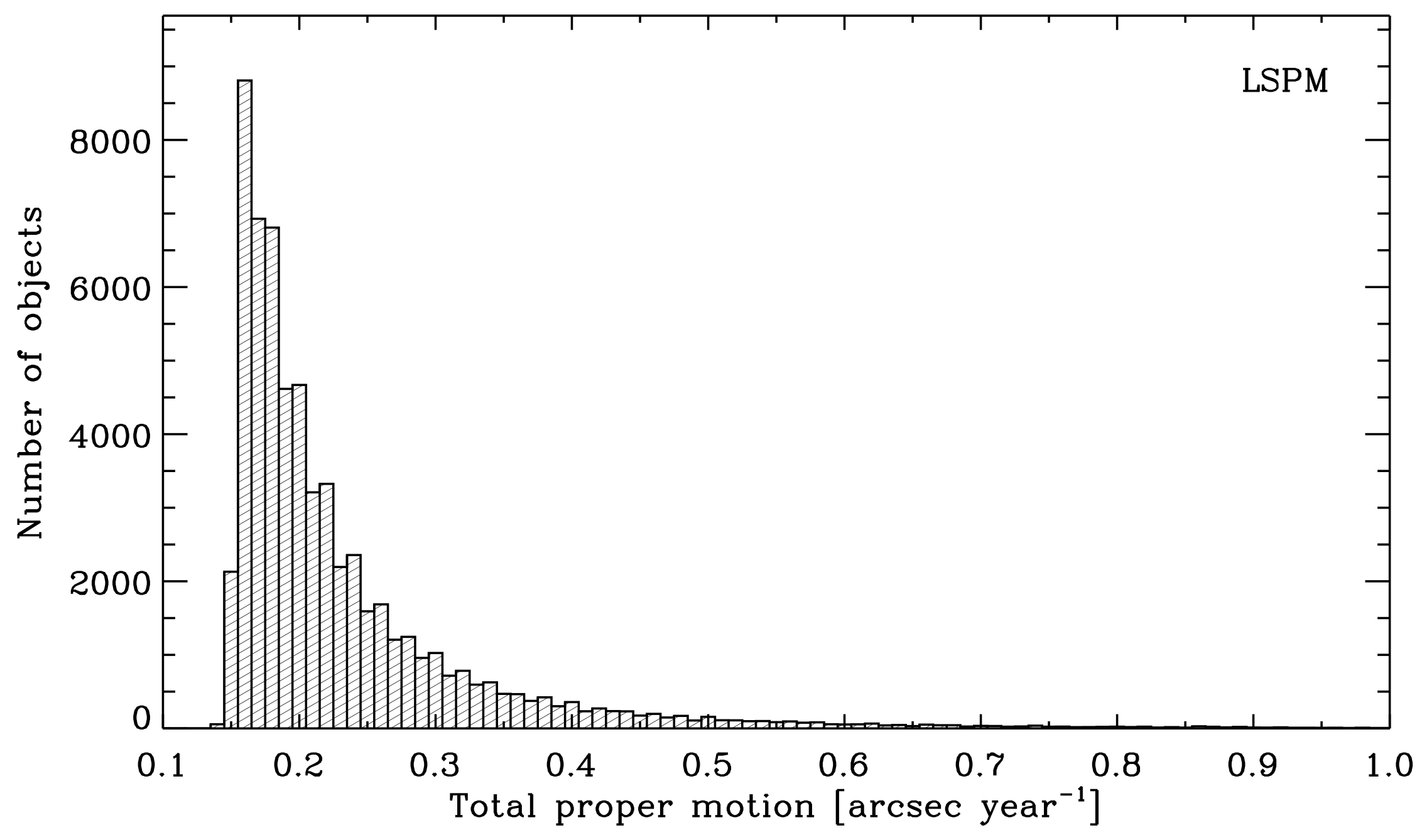}
\caption{LSPM proper motion distribution truncated at 1 arcsec yr$^{-1}$ \label{fig:lspm}}
\end{figure}
The difference in coordinate epochs can be of decades in the most unfortunate cases. In order to account for the difference in coordinate epochs and for the sake of completeness, when \emph{Gaia} proper motions are not available, we decided to increment the \emph{Gaia} position errors in order to
take into account not only the coordinates uncertainties, but also possible proper motions. 

For high proper motion stars with unknown proper motion in \emph{Gaia} DR1, we aim to obtain the completeness in the \emph{Neighbourhood} output, if not the correct match in the 
\emph{BestNeighbour} output: we may get the wrong counterpart (if another good neighbour is present), but the correct counterpart will probably be included in the \emph{Neighbourhood} table and can be recovered.
The effect of the unknown proper motions is potentially much larger than the position accuracy and depends strongly on the epoch difference
between  the \emph{Gaia} and the external catalogues sources. We thus consider the effect of the unknown proper motions as a bias rather than a systematic uncertainty. 
In this context, we consider as \emph{high} proper motion stars those objects whose motion in the sky, combined with the epoch difference between catalogues, 
can prevent a correct cross-match. The search radius is normally of the order of  a few arcsecs to account for the random and systematic errors on positions in the \emph{Gaia} and external catalogues. We thus consider as problematic high proper motion stars those   objects that, by combining the proper motion with the epoch 
difference, can travel a distance comparable with the search radius.

We adopted what we consider to be a reasonable solution: define a proper motion threshold common to all external catalogues. This threshold together with the epoch 
difference is used to define the initial search radius and to broaden the astrometric errors in the leading (i.e. \emph{Gaia}) catalogue (see Section~\ref{sec:details}). 
In order to define the proper motion threshold, we need to know the proper motion distribution of high proper motion stars. Figure~\ref{fig:lspm} shows the proper motion
distribution (truncated to 1 arcsec~yr$^{-1}$) of the LSPM high proper motion star sample. 
According to \citet{lspm}, the LSPM Catalog includes 61977 stars with total proper motion higher than 150 mas yr$^{-1}$ 
in the northern hemisphere and is complete to 99\% for stars at high galactic latitude ($|b|> 15$) and 90\% complete for stars at low galactic 
latitude ($|b|\leq 15$) at \emph{V}=19.0. 
Most of the stars ($\sim$74\%) in the LSPM Catalog have a proper motion smaller than 250 mas yr$^{-1}$, while $\sim$52\% of the stars have a proper 
motion lower than 200 mas yr$^{-1}$.
While ideally the proper motion threshold should be the maximum known proper motion of a real object, 
we fixed the proper motion threshold at 200 mas yr$^{-1}$ as a compromise between recovering large proper motion stars on the one hand and avoiding adding too many good neighbours and/or mismatches and preserving the performances of the XM calculations on the other.

Fixing a threshold for proper motions and using it to define the initial search radius (see Section~\ref{sec:details}) also influences  the cross-match of high proper motion stars with a measured \emph{Gaia} proper motion because the initial search area is defined around the \emph{Gaia} coordinates of a given object before applying the proper motion correction which depends on the epoch of the counterpart candidate. Fortunately the high proper motion stars with a measured \emph{Gaia} proper motion are a tiny fraction of the total: there are 6603 sources in TGAS with a total proper motion higher than 200~mas yr$^{-1}$. For GSC~2.3 we were able to match 6366 sources of the high proper motion subsample, while we recover 6182 of those sources in 2MASS  XM output. This problem will be solved in the XM algorithm planned for \emph{Gaia} DR2, and in the future high proper motion stars will be properly included in the XM output.
\subsection{Environment}
Given that the cross-match is both a source-to-source and a local problem, the definition of the best neighbour and the neighbour scoring should take into account  the surroundings of the \emph{Gaia} sources in the external catalogue. Therefore, the scientific information regarding the probability 
that a neighbour  is a good neighbour resides in the angular distance, but also in the local surface density of the external catalogues. 
The local density of the external catalogues is calculated on-the-fly by counting the number of external catalogue sources within a fixed radius circle centred  on each \emph{Gaia} source (see Sections~\ref{sec:details} and \ref{sec:extcat} for details).
This choice is not optimal, given the different densities between different catalogues and the large density variations with galactic coordinates within a given catalogue. It is, however, a trade-off between having a precise density (which requires a large number of stars and thus a large radius) and having 
a more accurate local density.  
By choosing a small radius (while keeping it much larger than the position errors with the exception 
of a few unfortunate cases), we obtain density estimates that are less precise, but more accurate.
It is certainly true that a large part of the sky contains small numbers of stars in small areas. However, this is exactly why   many of the objects in a survey are 
found in relatively dense areas. While a larger initial radius would be more appropriate for lower density regions, it would worsen 
the density determination when small-scale  density variations occur, especially in dense regions like the Galactic Plane.
Our choice is motivated by the fact that, even if we obtain a less precise density estimate in the {easiest} cases when the density is so low that there are very few candidates,  we do obtain both precise and accurate (i.e. very local) densities for the {difficult} dense fields. 
Finally, an advantage of the local density calculated on-the-fly is that it is measured around the \emph{Gaia} source position. The density is thus a characteristic of a given \emph{Gaia} source and common to all good neighbours evaluated. As such, it does not affect the best neighbour selection.

\section{\emph{Gaia} pre-computed cross-match: details\label{sec:details}}
The algorithm we prepared makes use of a plane-sweep technique which requires the catalogues to be sorted by declination, 
implies the definition of an active list of objects in the external catalogue for each \emph{Gaia} object, and allows  the input data to be read only once, thus making the XM computation faster (\citealt{Power2005}, \citealt{Abel2004},
\citealt{Devereux2004}, \citealt{Power2004}). 
In addition, we used a filter and refine technique: a {first filter} is defined by a large radius centred on a given \emph{Gaia} object and is used to select neighbours and
calculate the local surface density, while a {second filter}
is used to select good neighbours among neighbours. Good neighbours are thus filtered on an object-by-object basis.
The selection of the best neighbour among good neighbours is based on a figure of merit (`{score}' in the \emph{Neighbourhood} output table).

Great circle distances between \emph{Gaia} objects and counterpart candidates were evaluated using the haversine formula. We also made  some tests using
the special case for a sphere of the Vincenty formula obtaining identical results and very similar performances. 
A normal distribution for position errors is assumed and the position error ellipses are projected on the tangent plane. Even
if the position errors are not truly Gaussian, the probability density function is expected to be peaked
toward the mean within the error ellipse, and therefore a Gaussian is a reasonable approximation (see Sect.~3.2 for a discussion).

In the following the subscript \emph{G} stands for \emph{Gaia} and subscript \emph{E} stands for the external catalogue.
\subsection{Initial search radius (first filter)}
The initial search radius is defined around each \emph{Gaia} object as
   \begin{equation}
      R = \max(R_{\mathrm{density}}, R_{\mathrm{epochDiff}})
   ,\end{equation}
where $R_{\mathrm{density}}$ is the radius used to calculate the local surface density (60$\arcsec$) and
$R_{\mathrm{epochDiff}}$ is the radius needed to include in the XM output the stars with a proper motion up to the chosen threshold (200 mas yr$^{-1}$).
\noindent
The radius used to account for the epoch difference between catalogues is defined as
\begin{equation}
\label{eq:radepochdif}
R_{\mathrm{epochDiff}}= H_{\gamma} \cdot PosErr_{\mathrm{max}} + \left(\frac{PM_{\mathrm{ref}} \cdot EpochDiff_{\mathrm{max}}}{1000} \right)
,\end{equation}
where 
$H_{\gamma} = 5$ corresponds to a confidence level $\gamma = 0.9999994267$;
$PosErr_{\mathrm{max}}$  is the maximum of the combined position error;
$PM_{\mathrm{ref}}$ is the proper motion threshold; and
$EpochDiff_{\mathrm{max}}$ is the maximum reference epoch difference between \emph{Gaia} and the external catalogue.
\noindent
The maximum combined position error is defined as
\begin{equation}\begin{split}
PosErr_{\mathrm{max}} =~&\max[ \max(RAerr_{G}),\max(DECerr_{G})] + \\&\max[ \max(RAerr_{E}),\max(DECerr_{E}) ] 
\end{split},\end{equation}
\noindent
where  
$RAerr$  and $DECerr$  are respectively the uncertainties in Right Ascension and Declination.
\noindent
The maximum epoch difference between the two catalogues being matched is defined as
\begin{equation}\begin{split}
EpochDiff&_{\mathrm{max}} = \\& \max \Big[~|\max(refEpoch_{G})-\min(refEpoch_{E})|~,\\& |\min(refEpoch_{G})-\max(refEpoch_{E})|~ \Big]
\end{split}\end{equation}
with $R$  in arcsec, $PosErr_{\mathrm{max}}$ in arcsec, $PM_{\mathrm{ref}}$ in mas yr$^{-1}$, and $refEpoch$ in years.

The values of $EpochDiff_{\mathrm{max}}$ and $R_{\mathrm{epochDiff}}$ for all the external catalogues matched with \emph{Gaia} are listed in Table~\ref{table:ExtProp}. 
\subsection{Broadening of position errors }
While the \emph{Gaia} position errors for the TGAS subsample are propagated to the candidate counterpart epoch as described in Sect.~2.2,  for the 
majority of \emph{Gaia} sources we decided to account for the unknown proper motion systematic contribution. 
While how to define systematic uncertainties and estimate their magnitudes is open to debate, according to \cite{Pekka2003} the technique used should be consistent
with how the statistical uncertainties are defined since systematic and statistical uncertainties are then combined when the results are compared with theoretical predictions. As stated in the cited paper, {`a common technique for estimating the magnitude of systematic uncertainties is to determine the maximum variation in the measurement, $\Delta$, associated with the given source of systematic uncertainty. Arguments are then made to transform that into a measure that corresponds to a one standard deviation measure that one would associate with a Gaussian statistic, with typical conversions being $\Delta/2$ and $\Delta/\sqrt{12}$, the former being argued as a deliberate overestimate, and the latter being motivated by the assumption that the actual bias arising from the systematic uncertainty could be anywhere within the interval $\Delta$'.}
For the sake of completeness as discussed in Sects.~\ref{sec:gen},  ~3.1, and 3.4, we fixed a very high confidence level for the initial search radius and the statistical uncertainties, namely the 2D equivalent of $5\sigma$. We coherently decided to use the same proper motion threshold (200~mas yr$^{-1}$) used to 
define the initial search radius and the same high confidence level to define the unknown proper motion contribution. 
We increased the \emph{Gaia} position errors using the following equations:
\begin{equation}\begin{split}
&\sigma_{x_{G'}} = \sigma_{x_{G}} +  SysErr_{x} =  \sigma_{x_{G}} +  PM_{\mathrm{ref}} EpochDiff/5 \\&
\sigma_{y_{G'}} = \sigma_{y_{G}} +  SysErr_{y} =  \sigma_{y_{G}} +  PM_{\mathrm{ref}} EpochDiff/5
\end{split}
\end{equation}
For each of the external catalogues matched with \emph{Gaia}, Table~\ref{table:ExtProp} shows the maximum values ($SysErr_{\mathrm{max}}$) of the systematic contribution added. 
The actual size of the contribution varies with the exact epoch difference between a given \emph{Gaia} source and the external catalogue counterpart candidate 
being evaluated.
We recall here that, due to the presence of astrometric systematic uncertainties, position errors are not strictly Gaussian, even if the effect of unknown proper motions is not taken into account. Astrometric systematics are larger for the external catalogues than for \emph{Gaia}. In the case of the external catalogues, they are due to the process of linking the observation to the ICRS reference frame, for example. The number, brightness, and colour distributions of reference stars influence the astrometric solution and introduce systematics both globally and locally. This systematic effect is usually smaller than the effect of proper motions and epoch differences.
\subsection{Position error convolution ellipse}
For the definition of the convolution ellipse, we followed the approach described in
\citet{pineau} (see their Sect.~3 and Appendix~A for details).
The position error ellipses in equatorial coordinates are projected on the tangent plane centred on the \emph{Gaia} object position, with the x-axis in the direction towards the external catalogue counterpart candidate.
Position errors are respectively described as 2D Gaussians for \emph{Gaia}\footnote{In Equation~\ref{eq:err}, $\sigma_{x_{G'}}$ and $\sigma_{y_{G'}}$ stand for the broadened Gaia position errors defined in the previous section, with the exception of the TGAS subsample for which the errors were instead properly propagated using the known proper motion} and external catalogue objects   
\begin{equation}\begin{split}
&N_{G'}\left(x, y; \sigma^{2}_{x_{G'}}, \sigma^{2}_{y_{G'}} , \rho_{G'}\sigma_{x_{G'}}\sigma_{y_{G'}}\right) \\&
N_{E}\left(x - d, y; \sigma^{2}_{x_{E}}, \sigma^{2}_{y_{E}} , \rho_{E}\sigma_{x_{E}}\sigma_{y_{E}}\right)
\end{split}
\label{eq:err}
,\end{equation} 
where $d$ is the angular distance between the \emph{Gaia} object and the external catalogue counterpart candidate.
\noindent
The density of probability that the two sources are at the same location is given by the convolution product of the two distributions
\begin{equation}\label{eq:fxy}\begin{split}
N_{C}&\left(x, y; \sigma^{2}_{x_{C}}, \sigma^{2}_{y_{C}} , \rho_{C}\sigma_{x_{C}}\sigma_{y_{C}}\right) = f(x, y) = \frac{1}{2\pi\sigma_{x_{C}}\sigma_{y_{C}}\sqrt{1- \rho_{C}^{2} }}~\cdot \\& \exp \left[-\frac{1}{2(1- \rho_{C}^{2})} \left( \frac{x^{2}}{\sigma_{x_{C}}^2} + \frac{y^{2}}{\sigma_{y_{C}}^{2}} -
\frac{2 \rho_{C} x y }{\sigma_{x_{C}}\sigma_{y_{C}}} \right)\right]
\end{split},\end{equation}
where 
 \[
      \begin{array}{lp{0.8\linewidth}}
 \sigma^{2}_{x_{C}}= \sigma^{2}_{x_{G'}}+\sigma^{2}_{x_{E}}\\
 \sigma^{2}_{y_{C}}= \sigma^{2}_{y_{G'}}+\sigma^{2}_{y_{E}}\\
 \rho_{C}\sigma_{x_{C}}\sigma_{y_{C}} =  \rho_{G'}\sigma_{x_{G'}}\sigma_{y_{G'}} +  \rho_{E}\sigma_{x_{E}}\sigma_{y_{E}}\\
     \end{array}
\]
\noindent
By using the eigendecomposition of the variance-covariance matrix of $N_{C}$, defining $\sigma_{M}$ and $\sigma_{m}$ as the semi-major and semi-minor axis 
in the eigenvector frame $(x_1,y_1)$, changing to the polar coordinates ($r$,$\theta$) and integrating 
over $\theta$, the density of probability can be written as a Rayleigh distribution:
\begin{equation}
f(r) = r~\exp \left(~-\frac{1}{2}r^{2}~\right)
\end{equation}
where $r = \sqrt{\frac{x^{2}_{1} }{\sigma^{2}_{M}}+\frac{y^{2}_{1}}{\sigma^{2}_{m}} }$.
%
  \begin{figure*}
   \centering
   \includegraphics{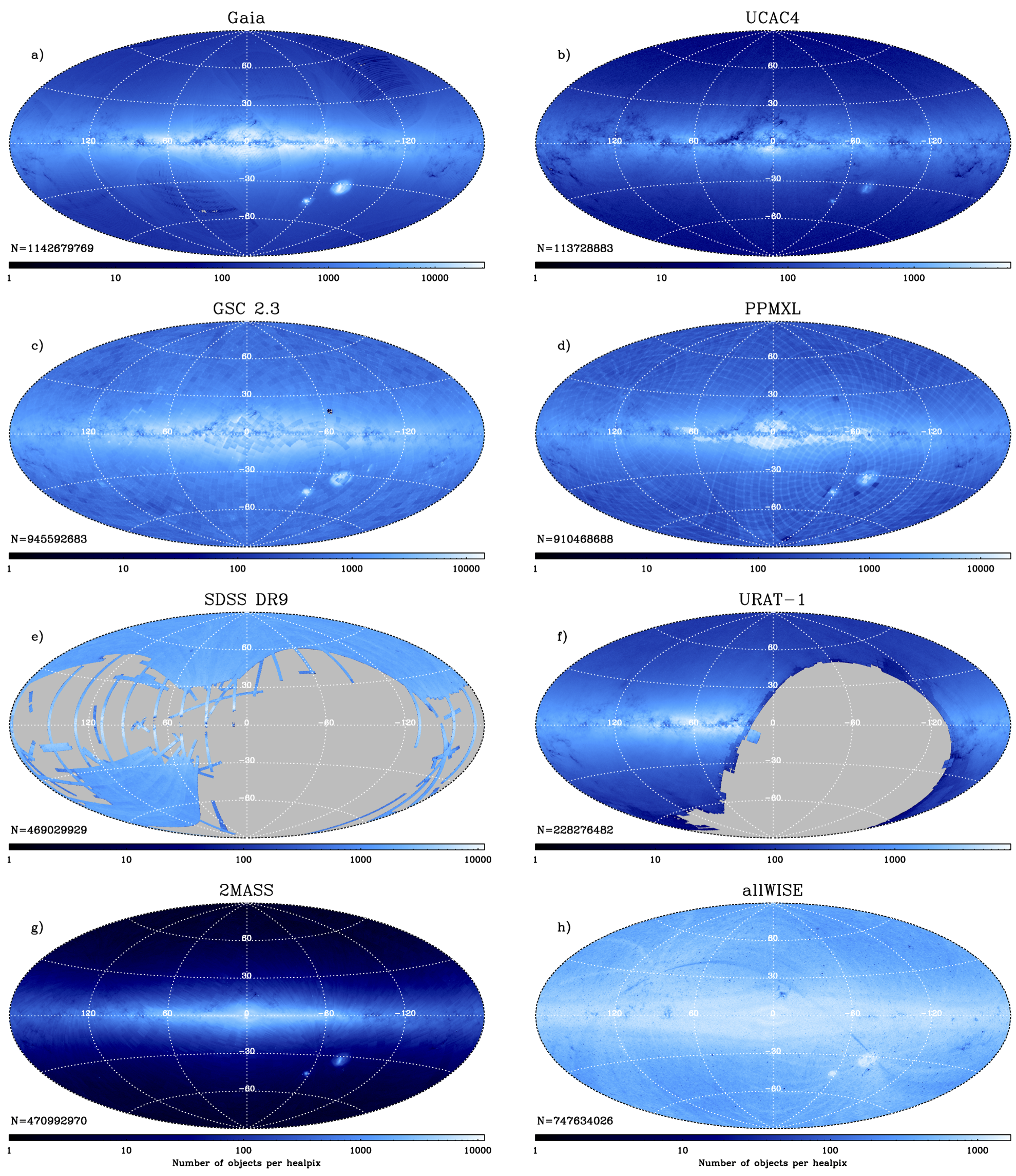}
   \caption{Surface density distribution for \emph{Gaia} and the external catalogues (see Section~\ref{sec:extcat}) obtained using a HEALPix 
                (\citealt{gorski}) tessellation with resolution $N_{\mathrm{side}}=2^{8}$. In grey are 
                indicated the areas not covered by 
                  the survey.}
              \label{figure:all}
    \end{figure*}
%
\subsection{Good neighbours' selection (second filter)}
The probability density function $f(x,y)$ defined in equation~(\ref{eq:fxy}) depends on $(x,y)$ only through
the exponent component
\begin{equation}\label{eq:kgamma} 
\frac{1}{1- \rho_{C}^{2}} \left( \frac{x^{2}}{\sigma_{x_{C}}^2} + \frac{y^{2}}{\sigma_{y_{C}}^{2}} -
\frac{2 \rho_{C} x y }{\sigma_{x_{C}}\sigma_{y_{C}}} \right)= K^{2}_{\gamma}
\end{equation}
or equivalently
\begin{equation}\label{eq:kgamma2} 
\begin{pmatrix}
x \\
y \\
\end{pmatrix}^t 
\Sigma^{-1} 
\begin{pmatrix}
x \\
y \\
\end{pmatrix} = K^{2}_{\gamma}
,\end{equation}
where $\Sigma$ is the covariance matrix (see also \citealt{pineau}) and
$K_{\gamma}$ is known as the Mahalanobis distance. Equation~(\ref{eq:kgamma}) defines the lines of constant probability
density; they are ellipses which define confidence regions (2D equivalent of confidence intervals), where $K^{2}_{\gamma}$ has
a $\chi^2$ distribution with 2 degrees of freedom. If we define $K_{\gamma}$ equal to a critical value of the $\chi^2$
distribution, the probability that $(x,y)$ will fall within the ellipse is equal to the confidence level $\gamma$
\begin{equation}
P\left\{  \frac{1}{1- \rho_{C}^{2}} \left( \frac{x^{2}}{\sigma_{x_{C}}^2} + \frac{y^{2}}{\sigma_{y_{C}}^{2}} -
\frac{2 \rho_{C} x y }{\sigma_{x_{C}}\sigma_{y_{C}}} \right) \leq  \chi^2_{2,\alpha}\right\} = 1- \alpha = \gamma
,\end{equation}
where $\alpha$ is the probability that $(x,y)$ will fall outside the ellipse.

\noindent
Good neighbours are defined as neighbours that fall within the ellipse defined by the confidence level $\gamma$:
\begin{equation}
\frac{1}{1- \rho_{C}^{2}} \left( \frac{x^{2}}{\sigma_{x_{C}}^2} + \frac{y^{2}}{\sigma_{y_{C}}^{2}} -
\frac{2 \rho_{C} x y }{\sigma_{x_{C}}\sigma_{y_{C}}} \right) \leq K^{2}_{\gamma} 
\end{equation}
Considering that external catalogue sources have coordinates $x=d$ and $y=0$,
the above equation becomes
\begin{equation}\label{eq:secondfilter}
\frac{d}{\sigma_{x_{C}} \sqrt{ 1 - \rho_{C}^{2}} } \leq K_{\gamma}\\
,\end{equation}
\noindent
where 
$d$ is the angular distance, 
$\sigma_{x_{C}}$ is the convolution ellipse error in the direction from \emph{Gaia} object to the possible counterpart, 
$\rho_{C}$ is the correlation between $\sigma_{x_{C}}$ and $\sigma_{y_{C}}$, and\
$K_{\gamma}$ depends on the confidence level $\gamma$:\\
if $\gamma$ = 0.9973002038  ,   $K_{\gamma}=\sqrt{11.8290}$;\\
if $\gamma$ = 0.9999366575 ,   $K_{\gamma}=\sqrt{19.3448}$; \\
if $\gamma$ = 0.9999994267 ,   $K_{\gamma}=\sqrt{27.6310}$. \\
The adopted value of $K_{\gamma}$ corresponds to a value of the confidence level $\gamma$ of 0.9999994267, which in 1D is equivalent to 5$\sigma$. 
The high confidence level was chosen in order to improve the completeness of the cross-match. It should be noted that the \emph{Neighbourhood} output table will contain only the good neighbours, which are not all the neighbours within a fixed radius, but all neighbours which are compatible within errors with the considered \emph{Gaia} source.
%
\subsection{Best neighbour selection: figure of merit}
The definition of the figure of merit is inspired by \citet{deruiter}, \citet{Wolstencroft}, \citet{Sutherland}, and \citet{pineau}. 
However, contrary to all of the above mentioned authors and consistently with the discussion in Sections~\ref{intro} and \ref{maginxm}, we did not add any {a priori} knowledge on the counterpart candidate's magnitude either as the number of possible counterparts in the magnitude bin of the candidate being considered or the number of possible counterparts brighter than the candidate being considered. In the specific scientific case addressed in this paper, we have no expectations on the brightness of the correct match.

The figure of merit (FoM) we used evaluates the ratio between two opposite models/hypotheses,
the counterpart candidate (i.e. the good neighbour) is a match or it is found by chance. The FoM
depends on the angular distance and the position errors (both used in the definition of the dimensionless variable $r$), on the epoch difference, and on the local surface density of the external catalogue. 

For each of the good neighbours, we compute the following FoM\footnote{The defined figure of merit is not a likelihood ratio because
 it is the ratio between two probabilities rather than between two likelihoods.}:
   \begin{equation}
      FoM(r) = \frac{dp(r|cp)}{dp(r|spur)}
      \label{eq:fom}
   \end{equation}
In equation (\ref{eq:fom}), $dp(r|cp)$ is the probability of finding a counterpart at a distance between r and r+dr :\\
\begin{equation}
dp(r|cp) = r~\exp \left(~-\frac{1}{2}r^{2}~\right)dr
\end{equation}
The probability of finding a spurious association is instead evaluated using the Poisson distribution, which is based on the assumption that celestial objects 
are locally randomly distributed and does not take into account the clustering of celestial objects. 
The Poisson probability of finding one or more objects by chance in an infinitesimal annulus area is
   \begin{equation}\begin{split}
    dp(r|spur) &= \sum_{k=1}^{\infty} \frac{s^{k}}{k!} \exp(-s) \, = \, \sum_{k=0}^{\infty} \frac{s^{k}}{k!} \exp(-s) - Poi(0,s) \\&
    = 1 - \exp(-s)  \approx  s \end{split}
   ,\end{equation}
where 
  \begin{equation}
   s = \rho \sigma_{M} \sigma_{m} \cdot dA \approx  \rho' \cdot 2 \pi r dr 
   \end{equation}
 is the number of sources within an infinitesimal annulus area $dA$, while the factor $ \sigma_{M} \sigma_{m}$ is needed to convert the measured density $\rho$ into the polar coordinates (as  was done in Sect.~3.3). The local surface density $\rho$ is defined by counting the number of objects within the initial search radius.
One of the reasons to prefer a more local surface density rather than a more precise one is that in order to increase precision, it is necessary to increase the size of the radius within which the density is evaluated making the assumption of a random distribution of sources less accurate.
 
 The figure of merit is thus
 \begin{equation}
 FoM(r) =  \frac{1}{2\pi\rho'}\exp\left({-\frac{1}{2}r^2}\right)
 \end{equation}
\noindent
In addition to making use of all the information available and not only the angular distance, the main advantage of the figure of merit with 
respect to a {nearest} neighbour is that it allows us to compare the {goodness} of best neighbours 
within a given catalogue. In the \emph{Neighbourhood} output table we listed the $\asinh$ of the figure of merit defined above:
 \begin{equation}
      score = \asinh(~FoM(r)~) = \asinh\left[\frac{1}{2\pi\rho'}\exp\left({-\frac{1}{2}r^2}\right)\right]
   \label{eq:score}
   \end{equation}
Since the FoM values cover a large range, applying an asinh make the output numbers more readable. It also has the advantage over 
the logarithm that it does not have a singularity at zero. 

The best neighbour is defined as the good neighbour with the highest score value, while mates are flagged and counted after the best neighbours have been evaluated for all \emph{Gaia} sources.
\begin{table}[b]
\caption{XM computation performance: computation time (see Section~\ref{sec:tech}).}
\begin{center}
\label{table:perf}
\begin{tabular}{l c}
\hline
Catalogue & Time \\
& (minutes)\\
\hline
UCAC4 &\phantom{222}39 \\
GSC2.3 &\phantom{22}239\\
PPMXL &\phantom{22}172 \\
SDSS DR9 &\phantom{222}56 \\
URAT-1 &\phantom{222}26 \\
2MASS PSC &\phantom{222}69 \\
allWISE &\phantom{22}450\\
\hline
\end{tabular}
\end{center}
\end{table}

\section{Technological implementation\label{sec:tech}}
The cross-match between catalogues which include hundreds of thousand of sources is a technological challenge, mainly because of the large number of angular distances required between pairs of sources in the two catalogues. Performance issues are even more critical when working within the 
framework of a large collaboration.
The XM implementation we developed  minimises the number of comparison by selecting a reasonable initial search radius (see Sect.~3.1), whose definition was a trade-off between completeness and performance, and calculating the angular distances for the second catalogue sources which would fall within the initial radius. In addition the Gaia catalogue was divided into several declination strips and the XM calculations for the different strips were run in parallel. 
\begin{table*}[t]
\small
\caption{Properties of  \emph{Gaia} and the external catalogues.}   
\centering          
\label{table:ExtProp}      
\begin{tabular}{ l  l  l  c  r  l  l  l  }     
\hline
Catalogue & $N$ Sources & $PosErr_{\mathrm{max}E}$\tablefootmark{a} & Effective Resolution &ICRS Offset & $EpochDiff_{\mathrm{max}}$&$SysErr_{\mathrm{max}}$\tablefootmark{b}&$R_{EpochDiff}$\tablefootmark{c}\\
                         &                       &\multicolumn{1}{c}{(arcsec)}  & (arcsec) &\multicolumn{1}{c}{(mas)}&\multicolumn{1}{c}{(yr)}  & (arcsec) &(arcsec) \\ \hline 
\emph{Gaia} DR1       &1\,142\,679\,769&\phantom{222}{0.1}                                 & 0.1\tablefootmark{d}   & ... &\phantom{2222} N/A & N/A & N/A \\
UCAC4           &113\,728\,883  &\phantom{222}2.279\tablefootmark{e}       &$\sim$2&    ...         &\phantom{222}16.88 & 0.65&\phantom{1}15.155\\ 
GSC 2.3         &945\,592\,683  &\phantom{222}1.6                                      & $\sim$4-5\tablefootmark{f}            &280       &\phantom{222}62.29&2.49&\phantom{1}20.958\\ 
PPMXL           &910\,468\,710  &\phantom{222}1.3421\tablefootmark{e}      & $\sim$4-5\tablefootmark{f}       &300       &\phantom{222}15.0 &0.60 &\phantom{1}10.211\\ 
SDSS DR9     &469\,029\,929  &\phantom{22}10.0                                      &  $\sim$1.4\tablefootmark{f}          &$<$100&\phantom{222}16.29 & 0.65&\phantom{1}53.758\\
URAT-1         &228\,276\,482  &\phantom{222}0.429                                   &  $\sim $5\tablefootmark{f}          &    ...           &\phantom{2222}2.689 &0.11&\phantom{18}3.183 \\ 
2MASS PSC  & 470\,992\,970  &\phantom{222}1.21                                   & 5         & 15        &\phantom{222}17.57 &0.70&\phantom{1}10.064\\ 
allWISE          &747\,634\,026   &\phantom{22}35.944                                & 6.1, 6.4, 6.5,12.0\tablefootmark{g}&    ...          &\phantom{2222}5.0 & 0.20&181.220\\ 
\hline          
\end{tabular}
\tablefoot{
   \tablefoottext{a}{$PosErr_{\mathrm{max}E} = \max [\max(RAerr_{E}),\max(DECerr_{E}) $], see Sect.~3.1.}
   \tablefoottext{b}{$SysErr_{\mathrm{max}} = PM_{\mathrm{ref}} EpochDiff_{max}/5$, see Sect.~3.2.}
   \tablefoottext{c}{$R_{EpochDiff}$ is defined in Sect.~3.1, Equation~\ref{eq:radepochdif}.}
   \tablefoottext{d}{The effective angular resolution on the sky of \emph{Gaia} DR1, in particular in dense areas, is not yet at this  
   expected level (\citealt{arenou}). }
   \tablefoottext{e}{The maximum of position error refers to the propagated errors at J2000.0.}
   \tablefoottext{f}{Effective resolution value is our best guess.}
   \tablefoottext{g}{Angular resolution in the four W1,W2,W3,W4 bands, respectively.}
   }
\end{table*}

In our implementation, the XM calculations are performed in RAM, and the input data for both the first and the second catalogues are read only once: for each Gaia source
an active list (which is a declination strip in the second catalogue) is defined and when passing to the following Gaia source the active list is updated, but not re-created.
The input data are organised in MariaDB 10.1 (a mysql fork) MyISAM tables, since MyISAM is a light MySQL storage engine suitable to fast reading. The output is written in the Percona XtraDB engine for MariaDB, which is an enhanced  version of the MySQL InnoDB engine. XtraDB is well suited for concurrent writing.
A large effort was put in the detailed configuration of MariaBD (and its engines)  in order to improve performance. While the input/output is supported by the MariaDB DBMS, the code is written in C language. All the calculations are performed in RAM by defining dedicated C data structures, which include the active lists and two different writing buffers for the \emph{BestNeighbour} and the \emph{Neighbourhood} outputs (see Section~\ref{sec:results}). The C data structures, as well as the number of Gaia strips and the size of the  external catalogues active lists, have been optimised for performance on the server we used (256GB RAM, two processors with 8 cores at 2.0~GHz with hyper-threading for a total of 32 CPUs and two 1.2TB SAS disks at 10K rpm). 
While the RAM was not an issue, the optimisation was a compromise between CPU usage and I/O limitations (actually the writing rather than the reading was a bottleneck). The performance depends on the characteristics of the external catalogues, mainly the stellar density. The best output performance was ~200.000 inserts/sec. 
Table~\ref{table:perf} includes the time needed to perform the XM computations for the different external catalogues. 
It should be noted that the reported times do not include the time needed to ingest and prepare the catalogues nor the time needed to run consistency tests on the results.
%
\section{External catalogue characteristics\label{sec:extcat}}
The following is the list of external catalogues cross-matched with \emph{Gaia} DR1 catalogue:
\begin{itemize}
\item{UCAC4 (\citealt{UCAC4});}
\item{GSC 2.3 (\citealt{GSC2.3});}
\item{PPMXL (\citealt{PPMX,PPMXL});}
\item{SDSS DR9 primary objects (\citealt {SDSS9,SDSS12});}
\item{URAT-1 (\citealt{URAT1}); }
\item{2MASS PSC (\citealt{2MASS});}
\item{allWISE (\citealt{WISE,allWISE});}
\end{itemize}
The main properties to consider when matching the external catalogues with \emph{Gaia} are 
\emph{a)}~angular resolution, \emph{b)}~astrometric accuracy, \emph{c)}~how the catalogue is tied to  the International Celestial Reference System (ICRS),
\emph{d)}~coordinates epoch, \emph{e)}~the need to propagate astrometric errors when the catalogue proper motions are available,\footnote{Positions are given at epoch J2000.0, but errors on positions are given at mean epoch.} and \emph{f)}~known issues and caveats.
It is also important to take into account how the external catalogue properties compare to \emph{Gaia} catalogue properties. 

Table~\ref{table:ExtProp}  lists the \emph{Gaia} and external catalogue properties relevant to the cross-match when they are available. 
Figure~\ref{figure:all} shows the sky coverage and the distribution of  the surface density for \emph{Gaia} and the external catalogues. 
The surface density was calculated by counting the number of sources in each pixel obtained using a  Hierarchical Equal 
Area and isoLatitude Pixelization (HEALPix; \citealt{gorski}) tessellation with resolution $N_{\mathrm{side}}=2^{8}$ which has 786\,432 pixels with 
a constant area of $\Omega\sim188.89~$arcmin$^{2}$.

The  external catalogue quantities used by our XM calculations are positions, position errors, position error correlation, and 
epochs. Different surveys may have  different definitions for  some of these quantities and/or use different units. For example, UCAC4 uses 
the south pole distance instead of declination, some surveys report the epoch in Julian years and some in MJD, and position errors are defined in different ways. 
The external catalogue input quantities were thus homogenised in order to simplify the XM calculations.

In the following we list some caveats and known issues that are relevant when computing the cross-match.
As stated in \citet{UCAC4}, in UCAC4 if the computed position error (at mean epoch) of a star exceeds 255 mas, it is set to 255 mas.
Similarly, according to the authors, the error in proper motion was truncated to 50 mas yr$^{-1}$, but respective stars were kept in UCAC4, if at least
two observations from different CCD observations were matched or the star is either in the 2MASS, SPM, or NPM data files.
Obviously, all large error objects need to be handled with caution, and some of these are simply non-existent.
Since the publication of UCAC4 in August 2012, the authors\footnote{\url{http://www.usno.navy.mil/USNO/astrometry/optical-IR-prod/ucac}} have suggested that the following corrections should be applied: identification of `streak objects'  and removal and data correction of a small number of high proper motion stars.

There are some 3\,350\,256 objects in GSC 2.3 with $RA$ and $DEC$ errors equal to 0, while it is mandatory to have errors on coordinates in order to run the cross-match. We decided not to exclude these objects and to assign to them the largest position error found in the catalogue.

There is a small number of objects (23\,945) in SDSS DR9 with large position error (greater than 10 arcsec and up to $\sim$14.36 degrees either in $RA$ or $DEC$). We decided to filter out these objects.
%
\begin{table}[t]
\small
\caption{\emph{BestNeighbour} output table content.}
\label{table:BestD}
\begin{tabular}{lp{0.5\linewidth}}
\hline
Field name & Short Description\\
\hline
SourceId &\emph{Gaia} source identifier\\
OriginalExtSourceId & Original external catalogue source identifier\\
AngularDistance &Haversine angular distance (arcsec)\\
NumberOfMates&Number of mates in \emph{Gaia} catalogue\\
NumberOfNeighbours&Number of good neighbours in external catalogue\\
BestNeighbourMultiplicity&Number of neighbours with same probability as best neighbour\\
ProperMotionFlag&Use of \emph{Gaia} proper motions (TGAS subsample)\\
\hline
\end{tabular}
\end{table}
\begin{table}[t]
\small
\caption{\emph{Neighbourhood} output table content.}
\label{table:NeigD}
\begin{tabular}{lp{0.5\linewidth}}
\hline
Field name & Short Description\\
\hline
SourceId &\emph{Gaia} source identifier\\
OriginalExtSourceId & Original external catalogue source identifier\\
AngularDistance &Haversine angular distance (arcsec)\\
Score & Figure of merit\\
ProperMotionFlag&Use of \emph{Gaia} proper motions (TGAS subsample)\\
\hline
\end{tabular}
\end{table}
\begin{table*}
\caption{\emph{BestNeighbour statistics}: Min/Max values of relevant output fields in \emph{BestNeighbour} tables. The fraction of \emph{Gaia} matched stars closer than 0.5 arcsec, those without mates and with a single neighbour, and  the number of \emph{Gaia} matched sources with no multiplicity are also listed. }
\centering
\label{table:Be}
\begin{tabular}{@{}lrccccccccc@{}}
\hline
Catalogue    & \multicolumn{1}{c}{Angular}  & \multicolumn{1}{c}{\% with } 
                       &\multicolumn{1}{c}{Number Of}  & \multicolumn{1}{c}{\% with Single}
                       &\multicolumn{1}{c}{Number } & \multicolumn{1}{c}{\% with} 
                       & \multicolumn{1}{c}{BestNeighbour}& \multicolumn{1}{c}{Sources}\\
                      & \multicolumn{1}{c}{Distance} &  \multicolumn{1}{c}{$d<0\farcs5$\tablefootmark{a}}
                      &\multicolumn{1}{c}{Neighbours} & \multicolumn{1}{c}{Neighbour}
                      & \multicolumn{1}{c}{Of Mates}      & \multicolumn{1}{c}{No Mates} 
                      &\multicolumn{1}{c}{Multiplicity}& \multicolumn{1}{c}{with $m>1$\tablefootmark{b}}\\ 
                      & \multicolumn{1}{c}{(arcsec)}  & \multicolumn{1}{c}{}
                      &\multicolumn{1}{c}{ }& \multicolumn{1}{c}{}     
                      & \multicolumn{1}{c}{ } & \multicolumn{1}{c}{}& \multicolumn{1}{c}{} & \multicolumn{1}{c}{}\\ \hline
                      & \multicolumn{1}{c}{max}  &\multicolumn{1}{c}{}  
                      & \multicolumn{1}{c}{max} &\multicolumn{1}{c}{}
                      & \multicolumn{1}{c}{max} &\multicolumn{1}{c}{}
                      & \multicolumn{1}{c}{max} &\multicolumn{1}{c}{}\\  \hline
UCAC4          &      9.87      &   86.95&       4    &  99.64  &   12     &   85.96   &  1  &0\\
GSC 2.3         &   17.86      &  60.11 &   18     &  90.04  &   18     &   70.72   &  16 &93964 \\
PPMXL          &     6.14       &  70.89 &   12     &  89.26  &     7       &  86.85   &   2  &6\\
SDSS DR9    &    51.88      &  98.54 &    8      &  98.60  &   61     &   99.48    &   1 & 0\\
URAT-1         &     2.12       &  99.80 &    2      &  99.99  &     3       &  99.75   &   1  &0\\
2MASS          &     6.82       &  85.32 &    5      &  98.92  &     6       &  88.03   &   2  &8\\
allWISE         &  181.11      &  77.60 &    3      &  99.99  &   20      & 99.15    &   1 & 0\\
\hline
\end{tabular}
\tablefoot{
\tablefoottext{a}{$d$=Angular Distance.} 
\tablefoottext{b}{$m$=BestNeighbour Multiplicity.} 
}
\end{table*}
\begin{table*}
\caption{\emph{Neighbourhood} statistics: Min/Max values of relevant output fields in \emph{Neighbourhood} tables.}
\centering
\label{table:Ne}
\begin{tabular}{@{}lclrll@{}}
\hline
Catalogue    & \multicolumn{1}{c}{Angular Distance}  & \multicolumn{2}{c}{Score} \\
                      & \multicolumn{1}{c}{(arcsec)} &  \multicolumn{2}{c}{ } \\  \hline         
                      & \multicolumn{1}{c}{max} 
                      & \multicolumn{1}{c}{min} & \multicolumn{1}{c}{max}  \\  \hline
UCAC4        & \phantom{18}9.87      &  0.000000002625   &  17.553515364854 \\
GSC2.3        & \phantom{1}18.54     &  0.000000000857   &  14.925308621133 \\
PPMXL        & \phantom{18}6.15      &  0.000000002021   &  15.234638147405 \\
SDSS DR9  & \phantom{1}52.40      &  0.000000000047   &  12.080360427928 \\
URAT-1        & \phantom{18}2.12      &  0.000000044589   &  18.779691208067 \\  
2MASS        & \phantom{18}6.82      &  0.000000004135   &  14.712484824539 \\
allWISE       &  181.11                       &  0.000000000514   &  12.744128884004 \\ \hline
\end{tabular}
\end{table*}
\section{Results\label{sec:results}}
The cross-match results are part of the official \emph{Gaia} DR1 release and are available at the ESAC Gaia Archive\footnote{\url{https://gea.esac.esa.int/archive/}} and at the SSDC Gaia Portal\footnote{\url{http://gaiaportal.asdc.asi.it/}}.

The XM output consists of two separate tables: \emph{BestNeighbour} includes the best matches (selected as the good neighbour with the highest 
value of the score), while \emph{Neighbourhood} includes all the good neighbours (selected using the second filter, see Equation~\ref{eq:secondfilter}). 
The XM output contents  are described in Tables~\ref{table:BestD} and \ref{table:NeigD}, respectively.
The BestNeighbourMultiplicity field in the \emph{BestNeighbour} output table addresses the binary stars and/or duplicates problem present in GSC 2.3, PPMXL, and 2MASS PSC. In these catalogues there is a fraction of source pairs with the same coordinates and position errors. 
Given that the astrometric properties are the same, the calculated XM score is also identical: the XM algorithm picks one of them, with no possibility to distinguish between the two. Fortunately,  these objects are quite rare,  as shown in the last column of Table  \ref{table:Be} where 
the number of sources with a bestNeighbourMultiplicity value greater than 1 is listed.

While it is not possible to discuss the correctness of the XM results on an object-by-object basis, it is possible to evaluate whether the general macroscopic results are as expected or if some features are present which could hint to a relevant fraction of mismatches. 

Tables~\ref{table:Be} and \ref{table:Ne} respectively show some statistics of \emph{BestNeighbour} and \emph{Neighbourhood} output tables for the different external 
catalogues matched with \emph{Gaia} DR1. The maximum value of angular distance and the fraction of matched pairs closer than $0\farcs5$ depend on the astrometric precision and on the epoch difference, so that their correlation with astrometric accuracy and systematics is less obvious. 
Table~\ref{table:Be} also shows that, even if in some cases the number of good neighbours for a given \emph{Gaia} source is large, the vast majority of \emph{Gaia} sources have a single neighbour in the external catalogue. 
The  lower fraction of Gaia sources matched with PPMXL and GSC2.3 sources which have only one good neighbour is most probably due to a fraction of duplicated sources in those catalogues at photographic plates borders which are visible in Figure~\ref{figure:all}.
In addition Table~\ref{table:Be} shows that the large majority of \emph{Gaia} objects do not have mates;  the \emph{Gaia} DR1 effective angular resolution on the sky is not yet at the expected level (\citealt{arenou}), which in turn is mainly due to heavy filtering in data reduction. The minimum and maximum score values  listed in Table~\ref{table:Ne} show that even matches with very low values of the figure of merit are kept in the XM output. The selection of good neighbours is based on the criterion defined in Equation~\ref{eq:secondfilter}, while no lower threshold for the figure of merit was fixed as it would be quite arbitrary. Matched pairs with a low score value should have a relatively large angular distance, large position errors, or be in fields with high stellar density.
\begin{table*}
\caption{External catalogue XM results: the number of objects compared with the number of matched sources, the fraction 
of matched \emph{Gaia} sources and the fraction of matched external catalogue sources. The number of sources in the \emph{Neighbourhood} tables is also listed.}
\centering
\label{table:Stat}
\begin{tabular}{lrrrrr}
\hline
Catalogue    & \multicolumn{1}{c}{Number of} & \multicolumn{1}{c}{Number of Best }  
                      &  \multicolumn{1}{c}{\% of \emph{Gaia}}  &  \multicolumn{1}{c}{\% of External cat}   
                      & \multicolumn{1}{c}{Number of}  \\
                      & \multicolumn{1}{c}{Sources}   & \multicolumn{1}{c}{matches\tablefootmark{a} }
                      &  \multicolumn{1}{c}{ sources matched\tablefootmark{a} }  &  \multicolumn{1}{c}{ sources matched\tablefootmark{a} }    
                      & \multicolumn{1}{c}{Neighbours}\\ \hline
UCAC4         &  113\,728\,883  &   117\,369\,911  & 10.3\phantom{$a$}  & 95.4&  117\,797\,078 \\
GSC2.3        &  945\,592\,683  &   844\,343\,562   & 73.9\phantom{$a$}  & 74.8&  937\,463\,454 \\
PPMXL         &  910\,468\,710  &   714\,367\,484   & 62.5\phantom{$a$} & 73.1&   801\,968\,492 \\
SDSS DR9  &  469\,029\,929  &    97\,018\,148    & 8.5\tablefootmark{b}     & 20.6 &    98\,411\,313  \\
URAT-1        &  228\,276\,482  &   209\,888\,464   & 18.4\tablefootmark{b}   & 91.8 &  209\,888\,621 \\  
2MASS         &  470\,992\,970  &   447\,946\,619  & 39.2\phantom{a}   & 89.3 &  452\,852\,361 \\
allWISE        &  747\,634\,026  &   311\,033\,691   & 27.2\phantom{a}  &41.4   & 311\,035\,922 \\ \hline
\end{tabular}
\tablefoot{
   \tablefoottext{a}{ `Number of Best matches' includes the \emph{mates}. This column and  `\% of \emph{Gaia}  sources 
   matched' indicate distinct \emph{Gaia} sources.  `\% of External cat sources matched' indicates the fraction of distinct external catalogue 
   sources matched.}
   \tablefoottext{b}{The percentage of matched \emph{Gaia} sources in this case does not take into account the external catalogue limited sky coverage (see Figure~\ref{figure:all}).}
   }
\end{table*}
Table~\ref{table:Stat}, Figure~\ref{FigResult1} and the histograms shown in Figure~\ref{FigHist} address the issue of completeness by showing how many \emph{Gaia} objects are matched and how many of the external catalogues sources are matched, and show respectively the sky and magnitude distribution of the matched sources. 
For example, the total number of sources in UCAC4 is around 10\% of the number of sources in \emph{Gaia} DR1, which is consistent with having 
$\sim$10\% of matched \emph{Gaia} sources and $\sim$95\% of UCAC4 sources matched. This is also consistent  with the flat sky distribution of  \emph{Gaia} versus UCAC4 best matches (shown in panels a and b of Figure~\ref{FigResult1}) and with the fact that matched and  unmatched UCAC4 sources are evenly distributed in magnitude.
On the contrary, GSC 2.3 and PPMXL reach fainter magnitudes outside the galactic plane but contain fewer objects closer to the galactic plane  
(probably due to their lower angular resolution compared to Gaia) as
shown in Figure~\ref{FigResult1}, panels c,d and e,f, respectively. This is reflected in the Table~\ref{table:Stat} results: a significant fraction of \emph{Gaia} sources are unmatched, as are  a significant fraction of GSC 2.3 and PPMXL sources.
The histogram in Figure~\ref{FigHist} shows that the GSC 2.3 and PPMXL sources with no \emph{Gaia} match are mainly faint sources ($Rf$ or $R2$ >19.0).
SDSS DR9 is much deeper in magnitude than \emph{Gaia} and this is clearly visible in the XM results shown in Figure~\ref{FigResult1}. 
The histogram in panel c of Figure~\ref{FigHist}  shows that the cross-match correctly selects the bright objects even if magnitudes are not directly used in the figure of merit definition.
In the case of the \emph{Gaia} versus allWISE cross-match, the results show that, as expected, the two surveys see quite a different sky.

The  comparison of the surface density distribution in the sky of all sources (see Figure~\ref{figure:all}) and of the matched sources (see Figure~\ref{FigResult1}), should help us understand whether some of the properties of the XM output are due to pairing failure or to features already present in the catalogues. For example, the sky area around 
$(l,b)=(-120\degr,+40\degr),$ where the \emph{Gaia} coverage is worse than  average, can also be clearly distinguished in the matched sources sky distributions.
\emph{Gaia} sources with a large number of mates (see Cols. 6 and 7 in Table~\ref{table:Be}) are not numerous and are usually in very dense fields or are matched with a source with large position errors in the external catalogue. Figure~\ref{manymates} shows a rather extreme example taken from the XM results for UCAC4. A single UCAC4 source (UCAC4~115-004819) is the best match for 13 different \emph{Gaia} objects.  The UCAC4 source is found in the very dense core of NGC~1718  star cluster ($RA$=$04^\mathrm{h}52^\mathrm{m}26\fs44, DEC$=$-67\degr03\arcmin08\farcs5$). 

Figure~\ref{fig:angdist} shows the angular distance distribution of the \emph{BestNeighbour} table for the different external catalogues. The top plots of each panel show the results of the published cross-match, while the bottom plots in each panel show the difference between the XM results calculated with and without the position error broadening. It is clear from the comparison between the top and bottom plots in each panel of 
Figure~\ref{fig:angdist} that without the position error broadening a large fraction of the matches are lost. On average, between  $\sim12$\% and  $\sim20$\% of the sources matched including the broadening are lost when original errors are used and epoch differences are ignored. The only exception is SDSS DR9 where a larger fraction of matches ($\sim65$\%) are lost.
In Appendix~\ref{sec:val} a validation test of the position error broadening approach using the TGAS subsample is described for 2MASS PSC, UCAC4, and GSC~2.3. The test shows that the effect of unknown proper motions and epoch differences is not negligible and that broadening the position errors leads to much more accurate and complete results than those obtained when ignoring this effect. 
   \begin{figure*}
   \centering
   \includegraphics{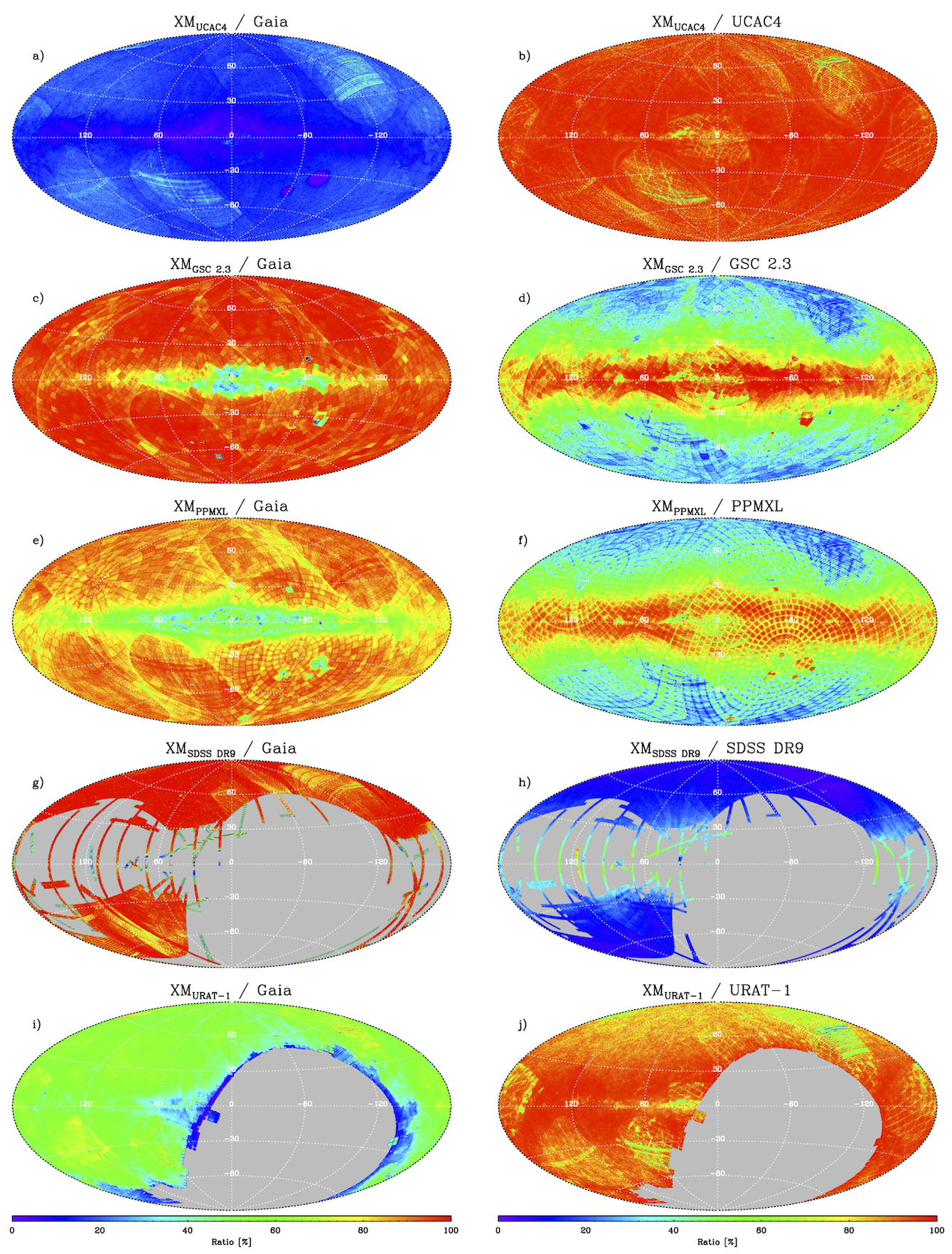}
   \caption{Surface density map for matched sources obtained using a HEALPix (\citealt{gorski}) tessellation with resolution $N_{\mathrm{side}}=2^{8}$. The left column figures show the fraction of \emph{Gaia} sources matched with an external catalogue, while the right
   column figures show the fraction of distinct external catalogue sources matched with \emph{Gaia}. }
              \label{FigResult1}
    \end{figure*}
  \setcounter{figure}{2}
   \begin{figure*}
   \centering
   \includegraphics{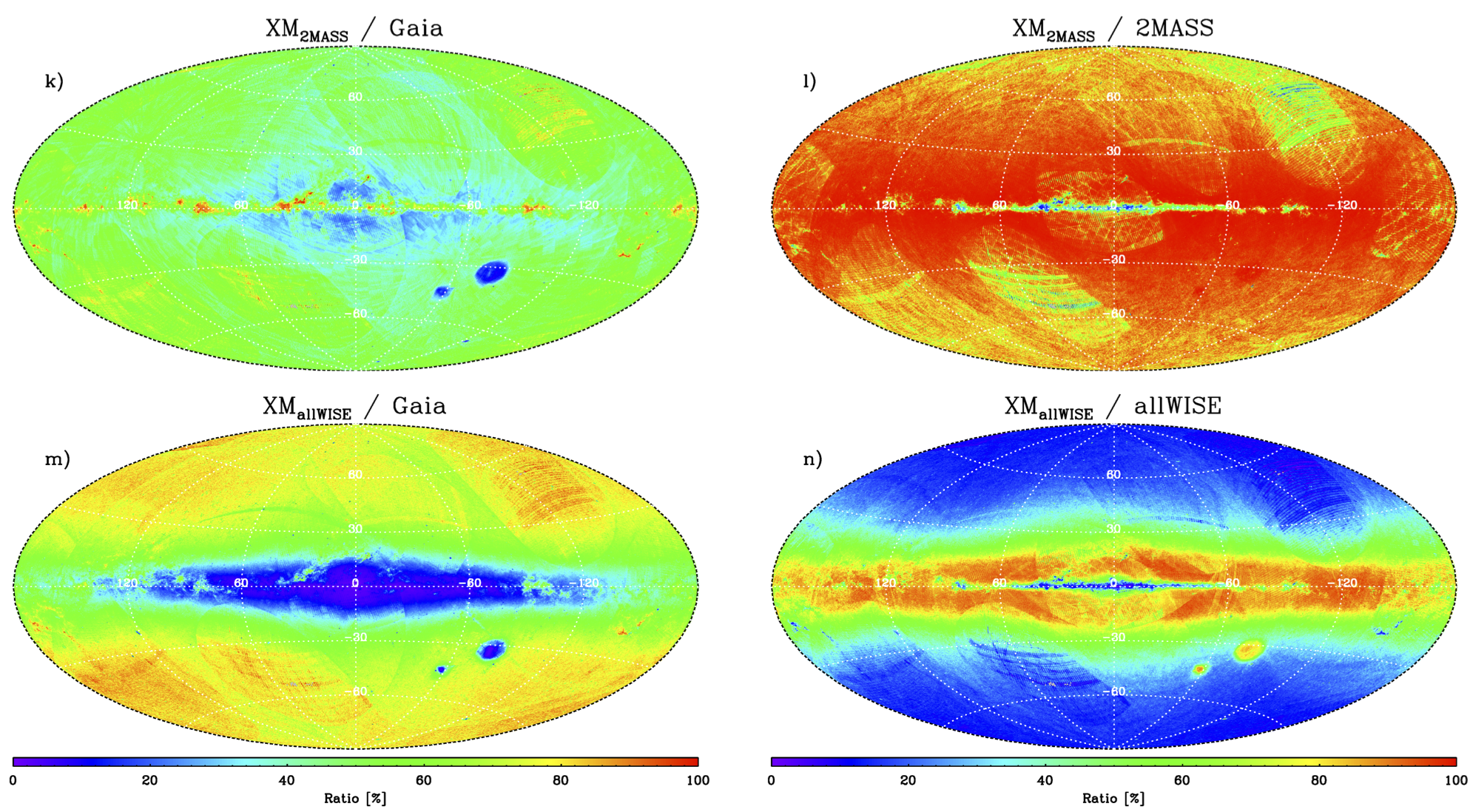}
   \caption{continued.}
    \end{figure*}

It should also be noted that the algorithm we used is not able to distinguish  between \emph{true} mates, which are objects resolved by \emph{Gaia} that are not resolved in the external survey, and \emph{false}/\emph{casual} mates. \emph{False} mates are a pair (or small group) of \emph{Gaia} objects with the same best neighbour in the external catalogue found at an angular distance which is larger than the effective resolution of the external catalogue. 
The discrimination between \emph{true} and \emph{false} mates thus depends on the angular resolution which is obviously much larger than position errors.
While  \emph{false} mates are present because  we chose a high confidence level ($5\sigma$)  and we incremented the position errors in order to account for unknown proper motions, it is not correct to consider \emph{false} all mates found at distances which are large compared with the position errors, but not compared to the angular resolution.
Matches which involve mates should be handled with particular care 
and a decision should be made about their being \emph{true} or \emph{false} on an object-by-object basis. In the case of \emph{true} mates, when combining \emph{Gaia} and an external catalogue photometry (for example in a colour magnitude diagram) the mates' fluxes should be added first.

The above analysis demonstrates that the XM results follow realistic expectations given the accuracy, precision, and diversity of input data sets, thus supporting the effectiveness of the adopted XM algorithm.
\section{Conclusions}
We developed a cross-match innovative in many respects and applied it in a consistent manner to a completely new survey of unprecedented astrometric accuracy such as \emph{Gaia} DR1.
The cross-match described in this paper is a large-scale XM which, quite uniquely, accounts for epoch differences and proper motions on an object-by-object basis. It uses an advanced algorithm based on a standard model of stellar motion when \emph{Gaia} proper motions are available, and instead  
adds a systematic contribution (which depends on a proper motion threshold and on the epoch difference) to the position errors  when \emph{Gaia} proper motions are not available. In addition, the position errors  are propagated to epoch J2000.0  for the surveys which list the coordinates at epoch J2000.0, but list position errors at a mean epoch (i.e. PPMXL and UCAC4). 
The adopted algorithm is also quite unique in that its definition of a many-to-one best match and the \emph{mate} concept is new and accounts for the \emph{Gaia} high 
angular resolution.
The definition of the output itself is original and much different from what is generally done. We tried to supply  scientists, in the output tables and in the analysis performed in this paper, with all the means to check the XM results and to understand whether this cross-match is appropriate for their scientific needs. They also have the completely new possibility of  overriding the best match choice we made by using the \emph{Neighbourhood} table and all the relevant quantities included there. For example, the angular distance and the figure of merit values could be complemented with {a priori} knowledge of counterpart magnitudes if a given scientific case benefits from it.

Since the XM algorithm described in this paper was developed for one very specific scientific case (i.e.  matching \emph{Gaia} data with large optical/IR surveys with an angular resolution lower than \emph{Gaia's}), it is not appropriate for matching \emph{Gaia} with the following:\\ 
\emph{a)} Catalogues with a comparable angular resolution (HST data for example). In this case the mates should not be present and a single \emph{Gaia} counterpart should be chosen for each external catalogue source;\\
\emph{b)} Catalogues obtained in wavelength regions different from optical/near-IR. In these cases the position accuracy and the density are very different from \emph{Gaia}'s. It is probably better to use the external catalogue as the leading catalogue and there are probably good reasons to use magnitude/colours or other a priori information in the best match choice; \\
\emph{c)} Lists of sparse objects. In these cases the completeness, the density, and the angular resolution of the sparse list are quite undefined. The best match is probably only one and should be chosen from the mates. The best match for a given Gaia source could well be a close source which is not included in the list. \\
\emph{d)} Catalogues of specific/peculiar objects rather than generic surveys. In these cases a priori information on the specific objects should be included in the best match selection criteria.
\begin{figure}
\centering
\includegraphics[width=80mm]{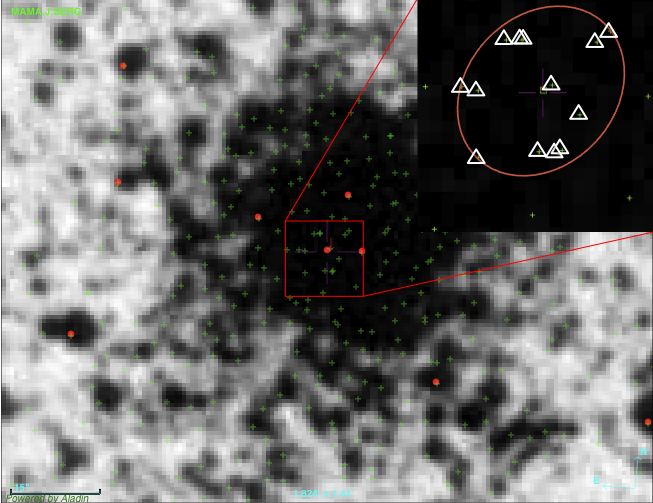}
\caption{Dense core of the NGC~1718 star cluster. Shown is the extreme case of UCAC4~115-004819, which is the best match for 13 different \emph{Gaia} objects. UCAC4 objects are indicated by filled red dots, while \emph{Gaia} sources by green crosses.
This figure was obtained using the CDS Aladin tool (\citealt{aladin1,aladin2}).}
\label{manymates}
\end{figure}
%
\begin{acknowledgements}
We would like to acknowledge the financial support of INAF (Istituto Nazionale di
Astrofisica), Osservatorio Astronomico di Roma,  ASI (Agenzia
Spaziale Italiana) under contract to INAF: ASI 2014-049-R.0 dedicated to ASDC.

This work has made use of data from the European Space Agency (ESA)
mission {\it Gaia} (\url{https://www.cosmos.esa.int/gaia}), processed by
the {\it Gaia} Data Processing and Analysis Consortium (DPAC,
\url{https://www.cosmos.esa.int/web/gaia/dpac/consortium}). Funding
for the DPAC has been provided by national institutions, in particular
the institutions participating in the {\it Gaia} Multilateral Agreement.

This research has made use of the  Aladin Sky Atlas developed at CDS, Strasbourg Observatory, France.

This research has made use of the VizieR catalogue access tool, CDS, Strasbourg, France. 
 
 We would like to thank R.~Smart, R.A.~Power, D.~Bastieri, G.~Fanari, and G.~Altavilla for very useful discussions, suggestions, help, and support. 
\end{acknowledgements}

%
   \begin{figure*}
   \centering
   \includegraphics[width=\linewidth]{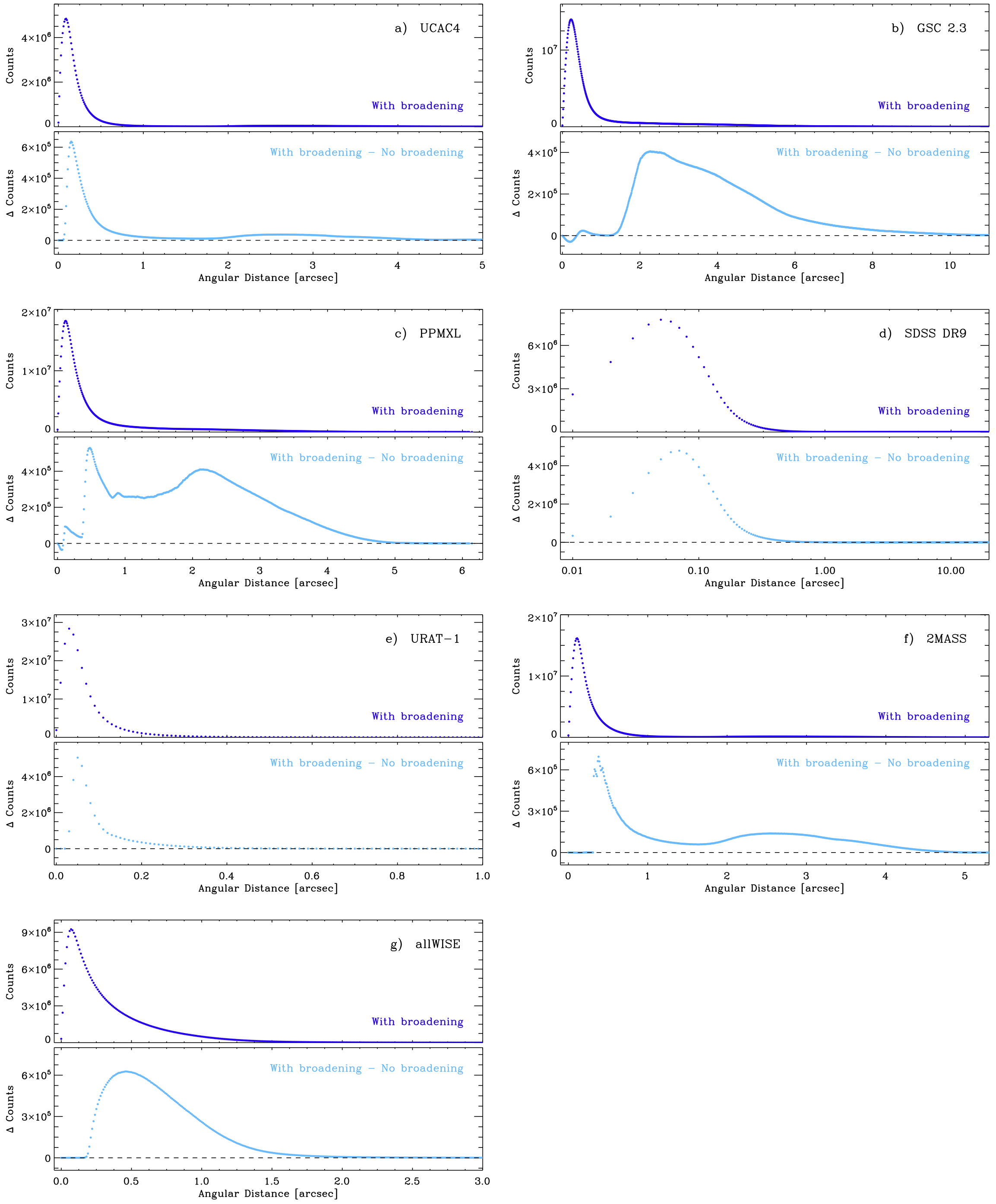}
   \caption{Angular distance distribution of the matched pairs in the \emph{BestNeighbour} table for the different external catalogues. For each catalogue   
   in the top plot of each panel the results are shown for the algorithm used for \emph{Gaia} DR1 cross-match (blue). The bottom plots of each 
   panel show instead the difference between the XM results calculated with and without the position error broadening (light blue).}
    \label{fig:angdist}
    \end{figure*}
   \begin{figure*}
   \centering
   \includegraphics[width=0.92\linewidth]{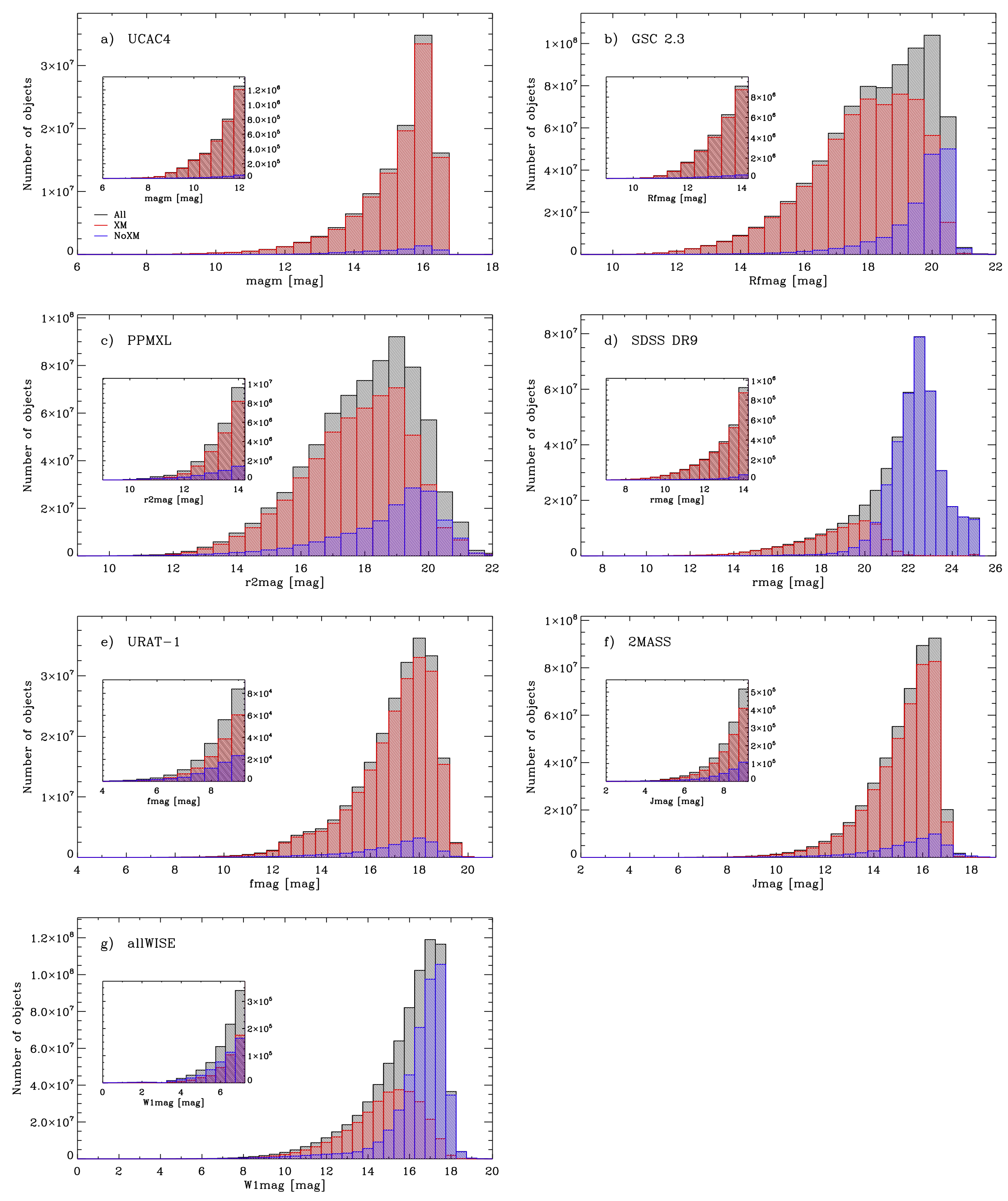}
   \caption{Magnitude distribution, in the most populated band, for the sources in the external catalogues. 
   In grey the catalogue distribution, in red the matched sources distribution, in blue the unmatched sources distribution.}
              \label{FigHist}%
    \end{figure*}
%
\bibliography{marrese}
\clearpage
\begin{appendix}
\section{Validation of position error broadening approach with TGAS\label{sec:val}}
We developed an XM algorithm which uses a new approach: the algorithm accounts for epoch differences and unknown proper motions rather than simply acknowledging  the impossibility of taking into consideration their effects. With the aim of comparing the results of these two different approaches,
\begin{figure*}
\centering
\includegraphics[width=0.75\linewidth]{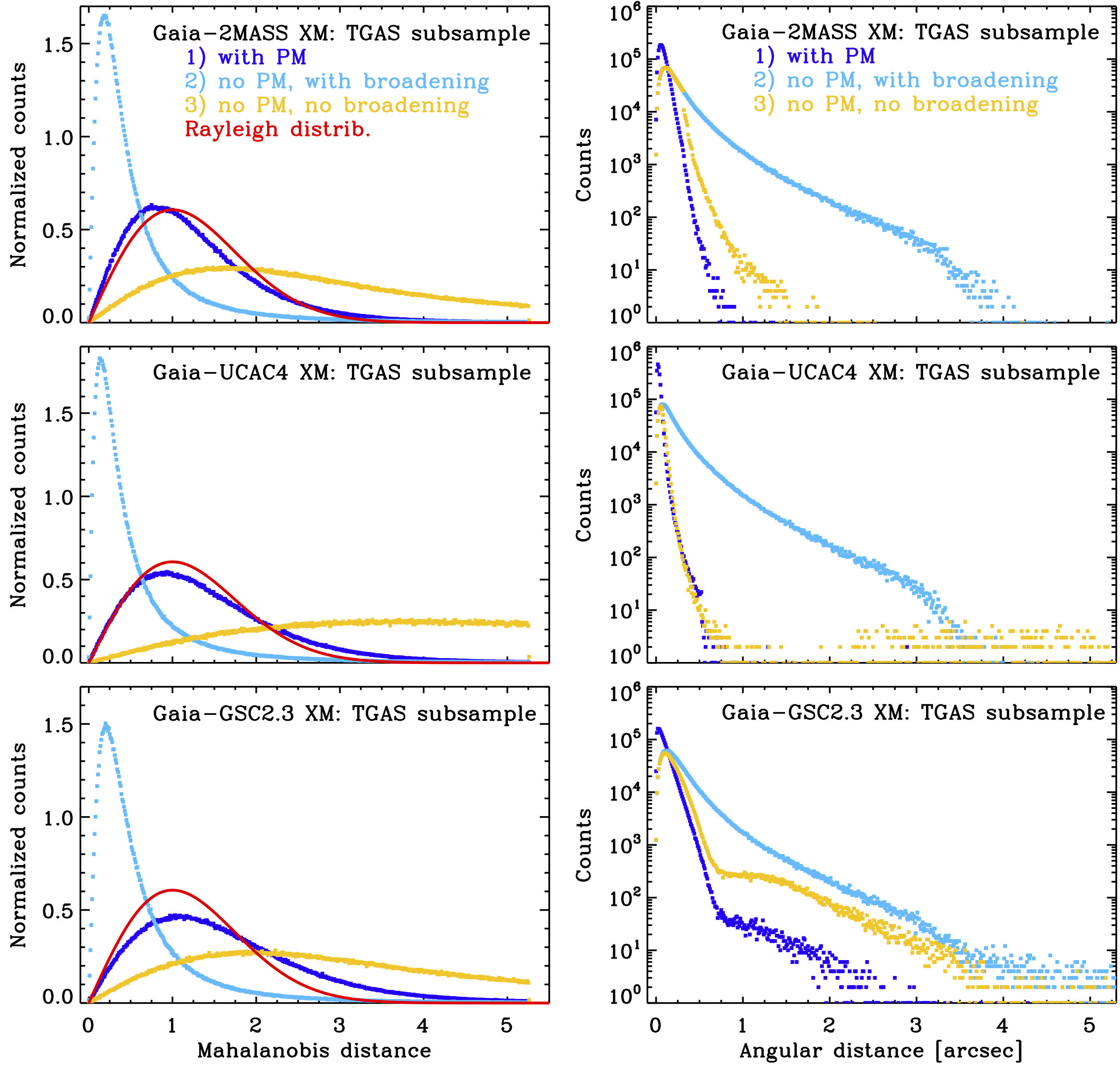}
\caption{Mahalanobis (i.e. normalised) distance and angular distance distributions for three different algorithms (position propagation using proper motions in blue,  position error broadening in light blue, and  original position errors in yellow) for the \emph{Gaia}-TGAS subsample matched with 2MASS PSC, UCAC4, and GSC 2.3.}
\label{fig:tgasall}
\end{figure*}
we describe here a simple test using the  \emph{Gaia}-TGAS subsample.
Three different XM algorithms were compared: 
\begin{enumerate}
\item{ properly propagate for proper motion positions and their errors (Sect.~2.2);}
\item{ ignore the proper motions and broaden the \emph{Gaia} position errors (Sects.~2.3 and 3.2);}
\item{ ignore the proper motions without applying any broadening to the \emph{Gaia} position errors.}
\end{enumerate}
It is important to note that using TGAS proper motions implies not only  moving the sources `closer' to their counterparts in the external catalogue, but also  propagating (i.e. broadening) the position errors.

For the three described algorithms, we performed the XM calculations between  the full \emph{Gaia} catalogue and 2MASS PSC, UCAC4\footnote{For UCAC4 the position errors are propagated to epoch J2000.0 in all cases.}, and GSC 2.3. We then extracted the \emph{Gaia}-TGAS
subsample from the corresponding \emph{BestNeighbour} tables. 

A comparison of the results for the TGAS subsample is shown in Figure~\ref{fig:tgasall}. The left panels show the Mahalanobis distance distributions in the three cases and compares them to the Rayleigh distribution. The plots show that in the first case (blue curves) the convolution of position errors is close to the theoretical expectations;  in the second case (light blue) it is, as expected, definitely overestimated; and in the third case (yellow)  it is clearly underestimated. 

Assuming that the results of the first algorithm are correct, the fraction of correct matches for the second and third algorithms are defined as the number of sources with the same best neighbour as obtained with the first algorithm. 
\begin{table*}
\caption{Comparison of the results of the three different XM algorithms for TGAS. The number of sources in TGAS is $2\,057\,050$.}
\centering
\label{table:tgas1}
\begin{tabular}{l|c|cc|cc}
\hline
Catalogue & N matches  & \multicolumn{2}{c|}{Fraction of matches} &  \multicolumn{2}{c}{Fraction of correct matches}\\
                  & algorithm 1 & algorithm 2 & algorithm 3 & algorithm 2 & algorithm 3 \\ \hline
2MASS PSC & $2\,045\,772$ &99.88&74.85& 99.39 & 74.76 \\
UCAC4 & $2\,042\,349$ &99.88&28.19& 99.20 & 28.08 \\
GSC 2.3\tablefootmark{a} & $2\,026\,532$ &100.72\tablefootmark{b}&63.79& 95.68 & 60.36 \\ \hline
\end{tabular}
\tablefoot{
   \tablefoottext{a}{See text for a discussion about GSC 2.3.}
    \tablefoottext{b}{In the case of GSC 2.3 we find a few more matches with broadening than with proper motion propagation.}
}   

\end{table*}
The results (summarised in Table~\ref{table:tgas1}) show that when a TGAS source is matched, all three algorithms produce, in the vast majority of cases, the same best match. However, the fraction of matches and the fraction of correct matches with respect to the 
first algorithm is always much larger when using the position error broadening (second algorithm) than when not broadening the position errors at all (third algorithm).

The right panels of Fig.~\ref{fig:tgasall} show that the same sources are, as expected, matched at larger angular distances using the second algorithm (light blue), since source positions are not propagated to the external counterpart epoch. When the third algorithm is used (yellow), only closer counterparts are found.
In the case of UCAC4 (which has much smaller position errors   than 2MASS or GSC 2.3) the third method fails to match a good fraction of the TGAS sources.

The high proper motion stars are not over-represented in TGAS because of the large time interval between Hipparcos/Tycho2 and  \emph{Gaia} observations. The TGAS proper motion distribution is in fact similar to the UCAC4 and PPMXL corresponding distributions.
However, given the smaller epoch difference between \emph{Gaia} and URAT-1 or allWISE (see column $EpochDiff_{\mathrm{max}}$ in Table~\ref{table:ExtProp}), the results of the second and third algorithms are expected to be less different in those cases (see also Figure~\ref{fig:angdist}).

This test shows that for the TGAS subsample the effect of unknown proper motions and epoch differences is not negligible and that
broadening the position errors leads to much more accurate and complete results than those obtained ignoring this effect.
  
We note that the TGAS subsample is not  fully representative of the entire \emph{Gaia} catalogue, given that the magnitude distribution is definitely different and bright sources are obviously  more common in TGAS. Likewise, the TGAS counterparts in the external catalogues are not fully representative of the entire surveys. There are  two main reasons for this: confusion and position error precision, which both depend on magnitude. 
Faint stars are difficult to detect around bright objects both for \emph{Gaia} and the external catalogues, and thus the confusion around TGAS sources (and their counterparts in the external catalogues) is lower than average. 

In addition, since bright (non-saturated) sources have normally more precise positions, this implies they are more difficult to match when astrometric systematics and proper motion effects are not taken into account (the third case described). This probably means that  when considering the full catalogues rather than the bright TGAS subsample, the inadequacy of the third algorithm is less severe. 

\begin{figure*}
\centering
\includegraphics[width=0.65\linewidth]{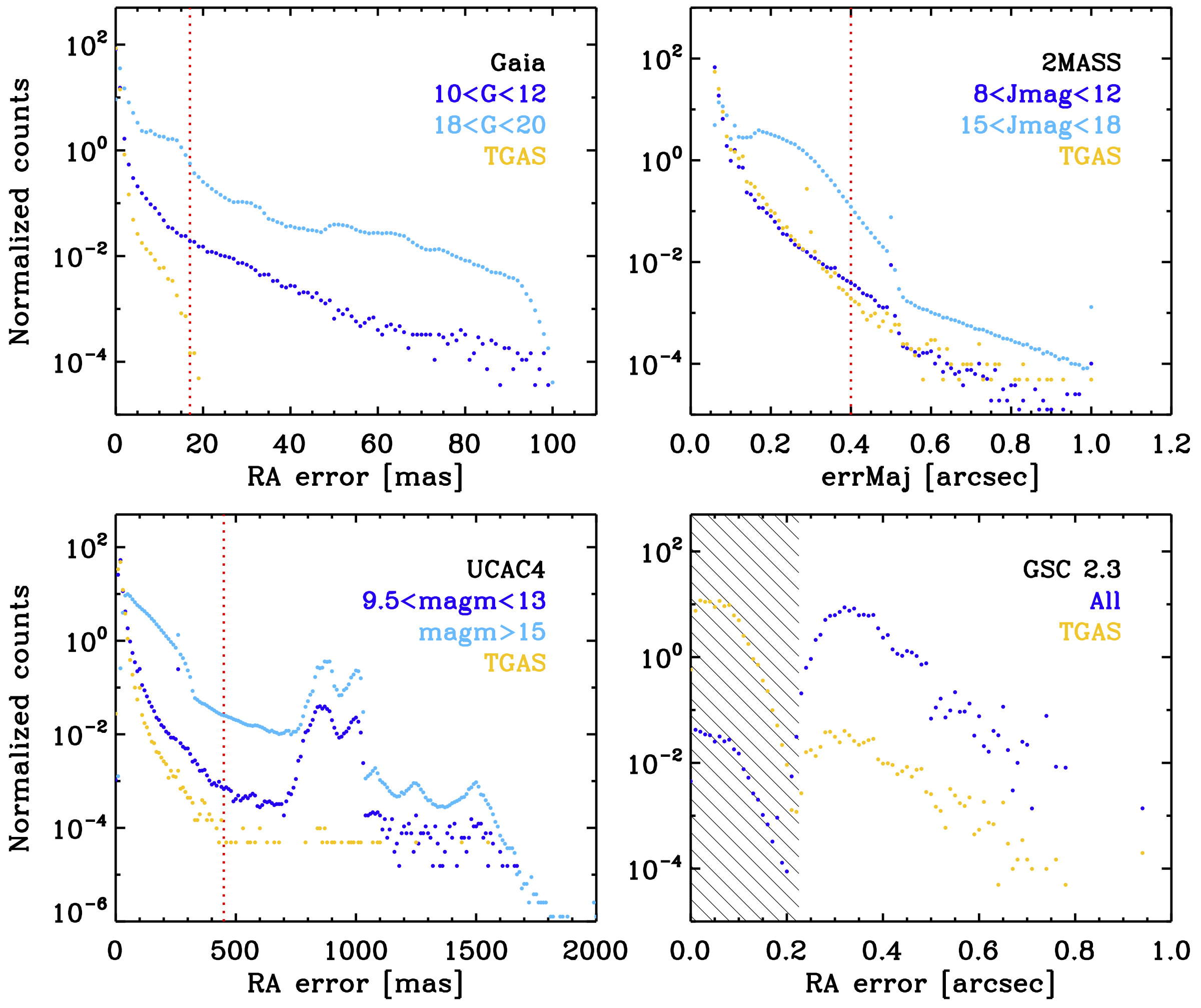}
\caption{RA error distribution for  \emph{Gaia}, 2MASS PSC, UCAC4, and GSC 2.3. The position error distribution is shown in light blue for the bulk of the sources in each catalogue, and in blue for all the sources in each catalogue in the same magnitude range as TGAS (for \emph{Gaia}) or TGAS counterparts (for the external catalogues). Each corresponding magnitude range is colour-coded and reported in the legend of each panel. In yellow is shown the distribution of the position errors for TGAS in the case of \emph{Gaia} or the TGAS counterparts in the external catalogues matched using the proper motions. For GSC 2.3 the blue dots indicate the full catalogue error distribution. The distribution of DEC error shows the same behaviour.}
\label{fig:tgasall2}
\end{figure*}

In order to assess to what extent the TGAS subsample is able to infer the validity of the position error broadening approach for the full catalogues, Table~\ref{table:tgas2} and Figure~\ref{fig:tgasall2} were prepared. 

For \emph{Gaia}, 2MASS and UCAC4,  Figure~\ref{fig:tgasall2} shows in light blue the position error distribution for the bulk of the sources in each catalogue (normally they are the faint magnitude end).
In yellow is shown the distribution of the position errors for TGAS (in the case of \emph{Gaia}) or the TGAS counterparts (in the external catalogues) matched using the proper motions. In blue is shown the same distribution for all the sources in each catalogue in the same magnitude range as the TGAS (for \emph{Gaia}) or TGAS counterparts (for the external catalogues). 
In Figure~\ref{fig:tgasall2} the dotted red vertical line indicates the position error threshold within which are contained  
$>99$\% of the TGAS or TGAS counterparts for each catalogue. Table~\ref{table:tgas2} summarises the fraction of sources with a position accuracy better than the threshold defined above.
Both Figure~\ref{fig:tgasall2} and  Table~\ref{table:tgas2} indicate that the TGAS subsample is a fair approximation of the full catalogues in terms of position error distribution.

The GSC 2.3 results, while very good for the TGAS subsample itself, cannot be used to infer the correctness of the position error broadening approach for the full catalogue. Unfortunately, the TGAS subsample counterparts are almost always found (see the hatched region in Figure~\ref{fig:tgasall2}) among the
bright stars which are saturated in the GSC 2.3 long exposure plates and were thus supplemented with data from Tycho-2 \citep{tycho2} and SKY2000 \citep{sky}, as reported in \citet{GSC2.3}. 

The effect of the difference in confusion between a bright sample and a full survey is to reduce the fraction of correct matches; however, our claim that broadening the position errors allows us to recover the matches in the \emph{Neighbourhood} table if not in the \emph{BestNeighbour} table does hold. This is obviously not true when position errors are not broadened.

A more complete validation of the position broadening approach will be possible with \emph{Gaia} DR2 data, when the vast majority of sources will have a published proper motion.
\begin{table*}
\caption{Comparison between TGAS (or TGAS counterparts) and full catalogue error distributions.}
\centering
\label{table:tgas2}
\begin{tabular}{lrrrr}
  \multicolumn{5}{c}{\emph{Gaia}} \\ \hline
Position precision&TGAS sources &All & Bright & Bulk \\
RA error <=17      & $\sim$100\% & 94.82\% &99.77 \% & 95.79\% \\ \hline 
&&&&\\ 
 \multicolumn{5}{c}{2MASS} \\ \hline
Position precision& TGAS counterparts & All & Bright & Bulk \\
errMaj<=400         & 99.99\% & 99.66\% & 99.97\% & 99.49\% \\ \hline 
&&&&\\ 
 \multicolumn{5}{c}{UCAC4} \\ \hline
Position precision & TGAS counterparts & All & Bright & Bulk \\
RAerror<=450      & $\sim$100\% & 96.45\% & 99.47\% & 95.32\% \\ \hline 
&&&&\\ 
 \multicolumn{5}{c}{GSC23} \\ \hline
Position precision & TGAS counterparts & All &  &  \\
RAerror<=225      & 99.48\% &0.35\% &  &  \\ \hline
\end{tabular}
\end{table*}
\end{appendix}


\end{document}